\documentstyle[epsf,emulateapj,apjfonts]{article}

\def\simlt{\mathrel{\hbox to 0pt{\lower 3.5pt\hbox{$\mathchar"218$}\hss}
      \raise 1.5pt\hbox{$\mathchar"13C$}}}
\def\simgt{\mathrel{\hbox to 0pt{\lower 3.5pt\hbox{$\mathchar"218$}\hss}
      \raise 1.5pt\hbox{$\mathchar"13E$}}}

\lefthead{FERRERAS \& SILK}
\righthead{PROBING THE EVOLUTION OF EARLY-TYPE GALAXIES 
THROUGH CHEMICAL ENRICHMENT}
\slugcomment{To be published in {\sl The Astrophysical Journal}.}

\begin{document}

\title{Probing the evolution of early-type cluster galaxies
	through chemical enrichment}
\author{Ignacio Ferreras \& Joseph Silk\altaffilmark{1}}
\affil{Nuclear \& Astrophysics Lab. 1 Keble Road, Oxford OX1 3RH, United Kingdom}
\altaffiltext{1}{Also at Department of Astronomy and Physics, University of
California, Berkeley CA 94720}
\authoremail{
1. ferreras@astro.ox.ac.uk\\
2. silk@astro.ox.ac.uk}

\vskip0.5truecm

\begin{abstract}
A simple chemical enrichment model for cluster early-type galaxies
is described in which the main mechanisms considered in the evolutionary model
are infall of primordial gas, outflows and a possible variation in the
star formation efficiency. We find that --- within the framework of
our models --- only outflows can generate a suitable range of metallicities 
needed in order to explain the color-magnitude relation. 
The chemical enrichment tracks can be combined with the latest population 
synthesis models from Bruzual \& Charlot (1999)
to simulate clusters over a wide redshift range, for a set of toy models
with different infall rates, star formation efficiencies and star formation
scenarios. The color-magnitude relation of local clusters is used as
a constraint, fixing the correlation between absolute luminosity and ejected
fraction of gas from outflows. It is found that the correlations between 
color or mass-to-light ratios and absolute luminosity are degenerate with 
respect to most of the input parameters. However, a significant change 
between monolithic and hierarchical models is predicted for 
redshifts $z\simgt 1$. The most important observable that differentiates 
between these alternative formation scenarios is the population of 
blue early-type galaxies which fall conspicuously blueward of the red 
envelope. The comparison between predicted and observed
mass-to-light ratios yield an approximate linear bias between total and
stellar masses: $M_{\rm Tot}\propto M_{\rm St}^{1.15\pm 0.08}$ 
in early-type galaxies. If we assume that outflows constitute the driving
mechanism for the colors observed in cluster early type galaxies, the metallicity
of the intracluster medium (ICM) can be linked to outflows: the color-magnitude 
constraint requires faint $M_V\sim -16$ galaxies to eject 85\% 
of their gas, which means that most of the metals in the ICM may 
have originated in these dwarf galaxies. No significant evolution 
is predicted, in agreement with X-ray observations 
(Mushotzky \& Loewenstein 1997\markcite{ml97}). Other mechanisms 
contributing to the enrichment of the ICM such as ejected material 
from mergers that formed the largest ellipticals should be translated 
into a decrease of the intracluster metallicity at $z\simgt 1-1.5$. 
Forthcoming observations from {\sl Chandra} and {\sl XMM} will help 
elucidate this point.
\end{abstract}


%%%%%%%%%%%%%%%%%%%%%%%%%%%%%%%%%%%%%%%%%%%%%%%%%%%%%%%%%%%%%%%%%%%%%%%%%%%%%

\section{Introduction}
Almost eighty years after the famous Great Debate between Curtis and Shapley
on the distance scale to spiral nebulae (Trimble 1995\markcite{tri95}), many 
issues in extragalactic astronomy have been solved. However, one of the most 
basic properties, namely the star formation history in galaxies, is still a 
matter of controversy. We only have a qualitative  description that 
separates early-type from late-type galaxies. The spectrophotometry-dynamics 
connection still waits to be established and the mechanism responsible for
tight correlations found in early-type galaxies such as the Fundamental Plane, 
the Color--Magnitude relation (CM) or the Magnesium--central velocity dispersion 
correlation still remains a mystery. While galaxies in rich clusters only
contribute around 5\% to the total number in the Universe 
(Bahcall 1996\markcite{ba96}), they live in a high density environment 
which provides a more advanced counterpart to field galaxy evolution 
(notwithstanding cluster specific issues such as interactions with the
hot intracluster medium). All galaxies in a given cluster 
lie approximately at the same distance, 
which makes the distribution in apparent and absolute luminosities the same, 
allowing observations of a reasonably large and unbiased sample. 
Even though we cannot infer from actual data that all cluster galaxies 
are coeval (i.e. a monolithic formation scenario), their evolution can 
also be traced --- on average --- by comparing clusters at different 
redshifts.

The study of the stellar populations in galaxies is complicated by the
presence of degeneracies, the most extreme of which is that between 
age and metallicity (Worthey 1994\markcite{wo94}). The effect of age or 
metallicity on many spectrophotometric observables is very similar:
for instance, a set of colors or spectral indices can be associated 
either with an old population of stars with a low metal content or with
a young population with higher metallicity. 
A recent analysis of a sample of clusters from
Stanford, Eisenhardt \& Dickinson (1998)\markcite{sed98} with redshifts
$0<z<1$ showed that the constraints from  multi-band 
(optical and near-infrared) color-magnitude 
relations still allow a wide range of formation redshifts, as low as
$z_F\sim 1$ for the faint end of the observed red envelope (Ferreras, Charlot
\& Silk 1999\markcite{fe99}). More targeted spectral indices such as 
Balmer absorption or the study of the mass-to-light ratios show better
age sensitivity, but the effect of metallicity combined with the
large uncertainties usually found in actual observations thwarts 
any reliable estimation of the age distribution of the stellar 
populations. The age-metallicity degeneracy thereby allows the
possibility of additional  mechanisms in order to explain the tight scaling relations 
found in early-type galaxies
other than a na\"\i ve assumption of a highly synchronous process 
of monolithic collapse in which the age spread of the stellar populations 
is very small. A highly stable process in the evolution of ellipticals 
could  result in a ``conspiracy'' that entangles age and metallicity in such a way that 
the stellar populations have a wide range of ages while keeping the 
correlations as tight as observed. Unfortunately, this very freedom
complicates estimates of the star formation history in galaxies,
especially early types, for which the major tracers of star formation
are absent, leaving only the imprint of the slowly-evolving population 
of intermediate and low mass stars in their spectral energy distribution.

A model-dependent way of breaking the age-metallicity degeneracy involves 
a chemical enrichment prescription, 
which combines the simplified treatment of galaxy evolution by the use
of a star formation rate and an adopted  distribution of stellar masses as they
are born (the initial mass function) along with our knowledge of stellar
evolution and nucleosynthesis. The most severe uncertainties come from the
yields, i.e. the mass fraction of stars transformed into metals for a given
mass, caused by the difficulties in modelling such stellar
evolution processes as stellar winds, mixing, the extremely high
sensitivity of nuclear cross sections to the temperature, and the
process of core collapse in high mass stars.
This makes any attempt at tracing the abundances of single elements
a qualitative rather than quantitative issue, and prompted our simplification
of the model, tracing only the net metallicity of the galaxy, i.e. the 
fraction of elements other than hydrogen or helium. In \S2 and 3
this model is described in detail as well as the simulation of the population
of early-type galaxies in a cluster. In \S4, 5 and 6 we explore the predicted
color-magnitude, $M/L$ vs mass relation and magnesium versus central velocity 
dispersion, respectively,  for a set of enrichment toy models as well as for 
a few star formation scenarios. Section 7 is devoted to describing a possible estimator 
of cluster evolution through the population of blue early-type galaxies. 
Section 8 analyzes the enrichment of the intracluster medium from galaxy 
ejecta, fundamental to the model presented in this paper. Finally, 
Section 9 discusses the principal conclusions.

%%%%%%%%%%%%%%%%%%%%%%%%%%%%%%%%%%%%%%%%%%%%%%%%%%%%%%%%%%%%%%%%%%%%%%%%%%%%%

\section{A Recipe for Chemical Enrichment}
In order to study the chemical enrichment of early-type galaxies in clusters
we will follow the formalism of Tinsley (1980)\markcite{ti80}, reducing the 
model to a set of a few parameters that govern the evolution.
A two-component system is considered, consisting of cold gas and stars.
Only the net metallicity will be traced, i.e. all elements heavier than helium
count in the same way. A more precise setup where the abundances of 
different elements are computed is beyond the scope of this project, which 
aims at finding the most basic mechanisms responsible for the observed 
range in metallicities that gives rise to the tight correlations obeyed 
by early-type galaxies, such as the color-magnitude relation. The mass in stars
($M_s(t)$) and in cold gas ($M_g(t)$) are normalized to the initial gas mass:
\begin{equation}
\mu_s(t)\equiv {M_s(t)\over M_g(t=0)}
\end{equation}
\begin{equation}
\mu_g(t)\equiv {M_g(t)\over M_g(t=0)}.
\end{equation}
We assume instantaneous mixing of the gas ejected by stars as well as 
instantaneous cooling of the hot gas component. The metallicity $Z(t)$
represents the average metal content of the different stellar populations
comprising the galaxy. The final spectral energy distribution comprises
the integration of stellar populations with different ages ``modulated''
by the star formation rate, and with an average metallicity given by
the chemical enrichment equations. Although
recent observations of moderate redshift clusters hint at high formation
redshifts $z_F\simgt 3$ (e.g. Ellis et al 1997\markcite{el97}; 
Stanford et al. 1998\markcite{sed98}), the age-metallicity
degeneracy is still capable of explaining the colors of the faintest
$z\sim 0$ galaxies (below $L_*$) with a formation redshift as low 
as $z_F\sim 1$ (Ferreras et al. 1999\markcite{fer99}). 
The basic physics related to chemical enrichment is reduced to a few 
mechanisms, namely:
\begin{itemize}
\item[$\bullet$] {\bf Infall:} Accretion of gas from outside is
   needed in order to explain the G-dwarf problem (Van den Berg 
   1962\markcite{vdb62}; Schmidt 1963\markcite{sch63}). Exponential
   infall at low metallicity is assumed: 
   $f(t) = \Theta(t-\tau_{lag}) A_{inf} 
	e^{-(t-\tau_{lag})/\tau_{inf}}$,
   where $\Theta(x)$ is the step function. The parameters 
   ($A_{inf}$,$\tau_{inf}$,$\tau_{lag}$)
   regulate the infall rate, timescale and delay, respectively. These
   infall parameters are constrained by the metallicity distribution of
   old disk stars (e.g. Prantzos \& Silk 1998)\markcite{ps98}.
\item[$\bullet$] {\bf Outflows:} Outflows triggered by supernovae
   explosions (SNe) constitute another important factor contributing to the final
   metallicity of the galaxy (Larson 1974\markcite{lar74}, Arimoto 
   \& Yoshii 1987\markcite{ari87}). Hence we define a parameter 
   $B_{out}$ which represents the fraction of gas that is ejected from  
   the galaxy. Eventually this parameter should be a function
   of the mass of the galaxy, whose potential well determines whether the
   SNe winds are strong enough to escape the pull of gravity. The outflow
   fraction is constrained by the deuterium abundance as well as the
   metallicity distribution of old disk stars 
   (Scully et al. 1997\markcite{scu97}; Cass\'e et al. 1998\markcite{cas98}).
\item[$\bullet$] {\bf Star Formation Efficiency:} A linear Schmidt law 
   (Schmidt 1963\markcite{sch63}) is assumed: 
	SFR$(t)\equiv\psi (t)=C_{eff}M_g(t)$, 
   where the parameter $C_{eff}$ gives the
   star formation efficiency, which should depend --- among other things ---
   on the mass of the galaxy as well as on whether a bursting episode is
   taking place. We will not worry here about a steeper, non-linear Schmidt law
   that may be required to account for the observed deuterium abundances
   (Cass\'e et al. 1998\markcite{cas98}), or for a possible enhancement 
   of the star formation rate with metallicity 
   (Talbot \& Arnett 1975\markcite{ta75}).
\end{itemize}

The aim of this chemical enrichment prescription is to break the degeneracy
between age and metallicity --- in a strongly model-dependent way --- so that
the large regions in (Age,Metallicity) space which are compatible with the
CM relation observed in early-type cluster galaxies can be
constrained (Ferreras et al. 1999\markcite{fer99}). The input 
parameters are only five: 
$(A_{inf},\tau_{inf},\tau_{lag},B_{out},C_{eff})$, 
with some additional initial
values which are: the initial metallicity and mass of stars and gas, the Initial
Mass Function (IMF) slopes and cutoffs, as well as the metallicity of the
infall gas. We will fix them hereafter to the following values:
\begin{itemize}
\item[$\bullet$] $\mu_s(t=0)=0$
\item[$\bullet$] $Z_g(t=0) = Z_f = 10^{-4}$
\item[$\bullet$] Scalo IMF (Scalo 1986\markcite{sca86}) with cutoffs at 0.1
	and 100$M_\odot$.
\end{itemize}

The equations can be separated into  one that follows  the mass
evolution and a second that  traces chemical enrichment.
\subsection{Mass Evolution}
The evolution of the mass in gas and stars is given by:
\begin{equation}
{d\mu_g\over dt} = (1-B_{out})E(t)-C_{eff}\mu_g(t)+
\Theta (t-\tau_{lag})A_{inf}e^{-(t-\tau_{lag})/\tau_{inf}}
\end{equation}
\begin{equation}
{d\mu_s\over dt} = C_{eff}\mu_g(t) - E(t)
\end{equation}
\begin{equation}
E(t) = \int_{m_t}^\infty dm\phi (m)(m-w_m)C_{eff}\mu_g(t-\tau_m).
\end{equation}
The integral $E(t)$ is the mass in gas ejected at time $t$ from stars which
have reached the end of their lifetimes. $\tau_m$ is the lifetime of a star
with mass $m$. The Instantaneous Recycling Approximation (IRA) is a simple 
relation for the lifetimes of stars, frequently used in analytic 
calculations (e.g. Tinsley 1980\markcite{ti80}),
which reduce $\tau_m$ to either zero or infinity depending on the mass
cut. This  considerably simplifies the calculation without having a huge
effect on the final metallicities, and is justified by the fact that
massive stars ($M\simgt 10M_\odot$) provide most of the metals and have 
lifetimes roughly 1000 times shorter than stars with solar masses, whose 
chemical contribution is negligible in comparison. However, we will solve the
equations numerically. Hence we will use actual lifetimes obtained from 
a broken power law fit to the data from Tinsley (1980)\markcite{ti80} and
Schaller et al (1992)\markcite{scha92}:
\begin{equation}
\left( {\tau_m\over { Gyr}} \right) = \left\{
\begin{array}{lr}
 9.694  \left( {M\over M_\odot}\right)^{-2.762} & M < 10M_\odot\\
 0.095  \left( {M\over M_\odot}\right)^{-0.764} & M > 10M_\odot\\
\end{array}
\right.
\end{equation}
$m_t$ is the turnoff mass, i.e. the mass of a main sequence star which reaches
the end of its lifetime at a time $t$. Finally, $w_m$ is the stellar remnant 
mass for a star with main sequence mass $m$:
\begin{equation}
\left( w_m\over M_\odot \right) = \left\{
\begin{array}{ll} 
 0.1(m/M_\odot )+0.45   & m/M_\odot < 10\\
 1.5                    & 10<(m/M_\odot)\leq 25\\
 0.61(m/M_\odot )-13.75 & m/M_\odot >25\\
\end{array}
\right.
\end{equation}
The mass for white dwarf remnants was taken from Iben \& Tutukov
(1984)\markcite{ib84}. The $1.5M_\odot$ remnant mass given for 
the intermediate range is the average mass of a neutron star
(e.g. Shapiro \& Teukolsky 1983),\markcite{shap83} whereas supernovae 
from heavier stars might give birth to black holes, locking more 
mass into remnants (Woosley \& Weaver 1995).\markcite{ww95}

\subsection{Chemical Enrichment}
The equations for the evolution of the metallicity of the gas and 
the stars are:
\begin{equation}
\begin{array}{ll}
d(Z_g\mu_g)/dt& =-C_{eff}Z_g(t)\mu_g(t)+\Theta (t-\tau_{lag})
Z_fA_{inf}e^{-(t-\tau_{lag})/\tau_{inf}}+\\
 & \\
 & +(1-B_{out})E_Z(t)\\
\end{array}
\end{equation}
\begin{equation}
\begin{array}{ll}
d(Z_s\mu_s)/dt& =C_{eff}Z_g(t)\mu_g(t)- \\
 & \\
 & -C_{eff}\int_{m_t}^\infty dm\phi (m)(m-w_m-mp_m)(Z_g\mu_g)(t-\tau_m)\\
\end{array}
\end{equation}
\begin{equation}
\begin{array}{ll}
E_Z(t)& =\int_{m_t}^\infty dm\phi (m)C_{eff}\big[ (m-w_m-mp_m)
	(Z_g\mu_g)(t-\tau_m)+ \\
 & \\
 & +mp_m\mu_g(t-\tau_m)\big],\\
\end{array}
\end{equation}
where $\phi(m)$ is the initial mass function (IMF). The fraction of 
a star of mass $m$ transformed into metals is given by $p_m$, and
is simplified by a triple power law (with cuts at $0.7$ and $1.0M_\odot$)
using the yields from Renzini \& Voli (1981)\markcite{rv81} and Marigo,
Bressan \& Chiosi (1996)\markcite{ma96} for Intermediate Mass Stars (IMS)
(i.e. $M\simlt 10M_\odot$). More massive stars undergo supernova explosions which
generate a large amount of metals. The yields in this mass range are taken from
Woosley \& Weaver (1995)\markcite{ww95} and Thielemann, Nomoto \& Hashimoto
(1996)\markcite{th96}. Rather important uncertainties arise because of the
complicated shock wave that is generated during core collapse. Usually a 
cutoff mass is assumed so that all matter above it is ejected into the
interstellar medium.  The differences in the yields between these two
groups are due to the different physical inputs assumed, mainly
the criterion for convection, the nuclear reaction rates, and the
determination of the mass cut

Type~Ia supernovae should also be considered as they contribute a large 
fraction of iron. These SNe are assumed to originate in a binary system in
which at least one of the stars is a white dwarf. The infall of gas from the
companion pushes the mass above the Chandrasekhar limit, triggering a 
deflagration with the subsequent disruption of the star. The thermonuclear
burning during this process generates roughly $0.7M_\odot$ of iron for a 
$1M_\odot$ $C+O$ white dwarf (W7 model of Thielemann, Nomoto \& Yokoi 1986
\markcite{th86}). We use the ratio of Type~Ia to Type~II (i.e. core collapse)
supernovae rates $N(Ia)/N(II)=0.12$ from Nomoto, Iwamoto \& Kishimoto (1997)
\markcite{no97}, in order to quantify the number of Type~Ia's that explode
at a given time. This estimation is based on a best fit to known solar
abundances, and so the ratio might be different for early-type galaxies.
However, this approximation is valid in the regime where early types 
have a high star formation rate, i.e. during the first few million years 
when the rate of type~II supernovae is significant. In the worst
case, this approximation will translate into a systematic underestimate 
of the iron yield whose effect will be diminished by the imposed 
color-magnitude constraint.
The equations are integrated using a fifth-order Runge-Kutta method with 
adaptive stepsize control given by the Cash-Karp prescription (Press 
et al. 1992\markcite{pre92}). The integration does not find any
special numerical hurdle as all functions entering the differential 
equations are smooth power laws. Only some
rescaling was necessary, changing the step scale from linear to 
logarithmic when computing the mass integrals in order to achieve 
convergence with a small number of steps.

The evolution of the metallicity for a range in all five parameters 
$(A_{inf},\tau_{inf},\tau_{lag},B_{out},C_{eff})$ 
is shown in figure~1, where the (Age,Metallicity) regions
for the brightest ($2L_*$) and faintest ($L_*/4$) bins of the early-type
cluster population of a cluster at $z\sim 0.5$ are overlaid 
(Ferreras et al. 1999\markcite{fe99}). 
The main conclusion that can be drawn
from the chemical enrichment tracks is that galaxy outflows 
($B_{out}$) constitute the principal driving mechanism for
generating a suitable range of metallicities 
that can explain the tight correlations in early-type galaxies such as
the Color-Magnitude relation. With respect to infall of gas: neither 
the rate ($A_{inf}$), timescale ($\tau_{inf}$) nor delay 
($\tau_{lag}$) can change the metallicities significantly unless 
pre-enrichment of the infall gas 
($Z_f >> 10^{-4}$) or a time dependence $Z_f(t)$ for its metal content are
assumed. Needless to say, the final mass in stars and gas can be greatly 
changed by infall, but the metallicity is roughly the same. 
A range of star formation efficiencies ($C_{eff}$) will only change
the timescale to reach some asymptotic final metallicity which is the same
regardless of the value of $C_{eff}$. 

%%%%%%%%%%%%%%%%%%%%%%%%%%%%%%%%%%%%%%%%%%%%%%%%%%%%%%%%%%%%%
%%%%%%%%%%%%%%%%%%%%%%%%   FIG. 1   %%%%%%%%%%%%%%%%%%%%%%%%%
%%%%%%%%%%%%%%%%%%%%%%%%%%%%%%%%%%%%%%%%%%%%%%%%%%%%%%%%%%%%%

\centerline{\null}
\vskip3.4truein
\includegraphics{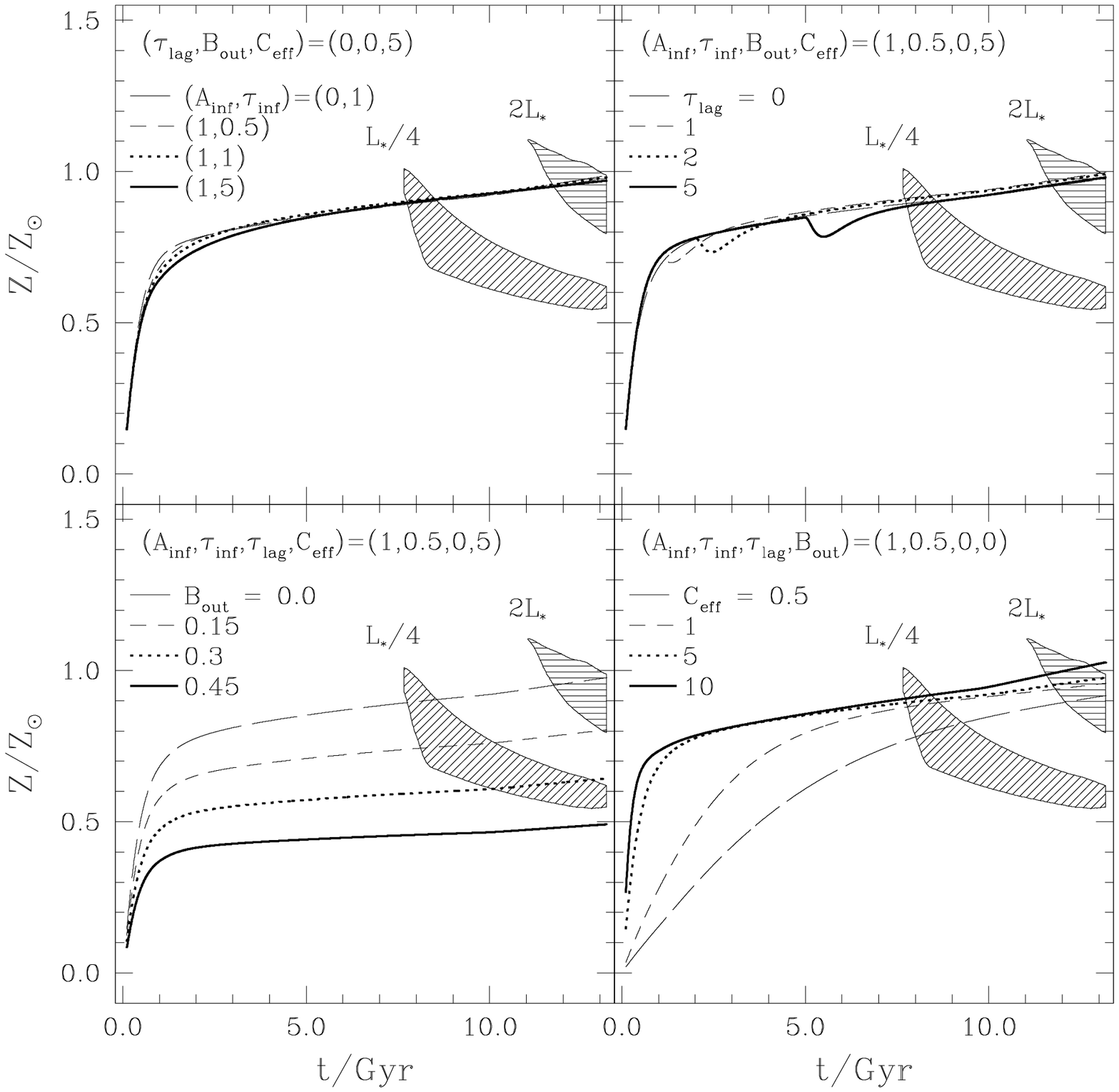}
\figcaption[fs_f1.eps]{Chemical enrichment tracks overlaid on
the age and metallicity regions for two luminosity bins --- 
corresponding to $2L_*$ and $L_*/4$ --- for cluster Cl0016+16 
($z=0.545$) using Color-Magnitude constraints (Ferreras et al. 1999).
Different ranges of infall are shown in the upper panels: Infall rate 
and timescale ($A_{inf}$ and $\tau_{inf}$, respectively, {\sl left}), 
and Infall delay ($\tau_{lag}$ in Gyr, {\sl right}). The bottom right 
panel shows the chemical 
enrichment tracks for a range of star formation efficiencies ($C_{eff}$),
whereas the bottom left panel plots the evolution of metallicity for a
few ejected fractions ($B_{out}$). Only this one is capable of
generating a suitable range of metallicities that yields the color
range of the red envelope.\label{f1}}
\vskip+0.2truein

%%%%%%%%%%%%%%%%%%%%%%%%%%%%%%%%%%%%%%%%%%%%%%%%%%%%%%%%%%%%%

\noindent
This final metallicity depends on more
basic factors such as the shape of the IMF or the yields from massive stars.
Hence, if we want to avoid any unphysical fine tuning in the galaxy formation
process in order to explain the CM relation, we conclude that
outflows from SNe-triggered winds are the main cause of the observed range of
metallicities in early-type cluster galaxies. Since the ejected fraction
parameter $B_{out}$ is mainly related to the mass of the galaxy, we could even 
extend this statement to field early-type galaxies. 

Figure~2 shows a mapping of
the enrichment tracks shown in figure~1 into a color-age diagram, where
restframe $U-V$ color is plotted versus formation redshift 
({\sl bottom panels}). Throughout this paper, the translation between 
ages and redshifts is done using an open cosmology with 
$\Omega_0=0.3$ and $H_0=60$ km s$^{-1}$ Mpc$^{-1}$. 
The shaded area encompasses the range of $U-V$ colors observed in 
the Coma cluster (Bower, Lucey \& Ellis 1992\markcite{ble92}, BLE92). 
Even though figure~2 shows that a range in infall parameters or 
star formation efficiency could explain the observed color range, 
one should take into account the star formation rate: the top panels
trace the star formation rate through the equivalent width
of H$\alpha$ using the standard prescription (e.g. Kennicutt 1998)
\markcite{ken98}. The choices of parameters that yield the bluest
colors of observed early-type cluster galaxies ($U-V\sim 1.2$) imply
a non-negligible H$\alpha$ emission line for a wide range of
formation redshifts and thus must be ruled out.
The dotted horizontal line sets an observational upper limit 
EW(H$\alpha$)$<$0.5\AA\  below which all realistic models should lie.
Hence, models with large amounts of infall, long infall delays or 
low star formation efficiencies are not considered.

%%%%%%%%%%%%%%%%%%%%%%%%%%%%%%%%%%%%%%%%%%%%%%%%%%%%%%%%%%%%%
%%%%%%%%%%%%%%%%%%%%%%%%   FIG. 2   %%%%%%%%%%%%%%%%%%%%%%%%%
%%%%%%%%%%%%%%%%%%%%%%%%%%%%%%%%%%%%%%%%%%%%%%%%%%%%%%%%%%%%%

\centerline{\null}
\vskip3.4truein
\includegraphics{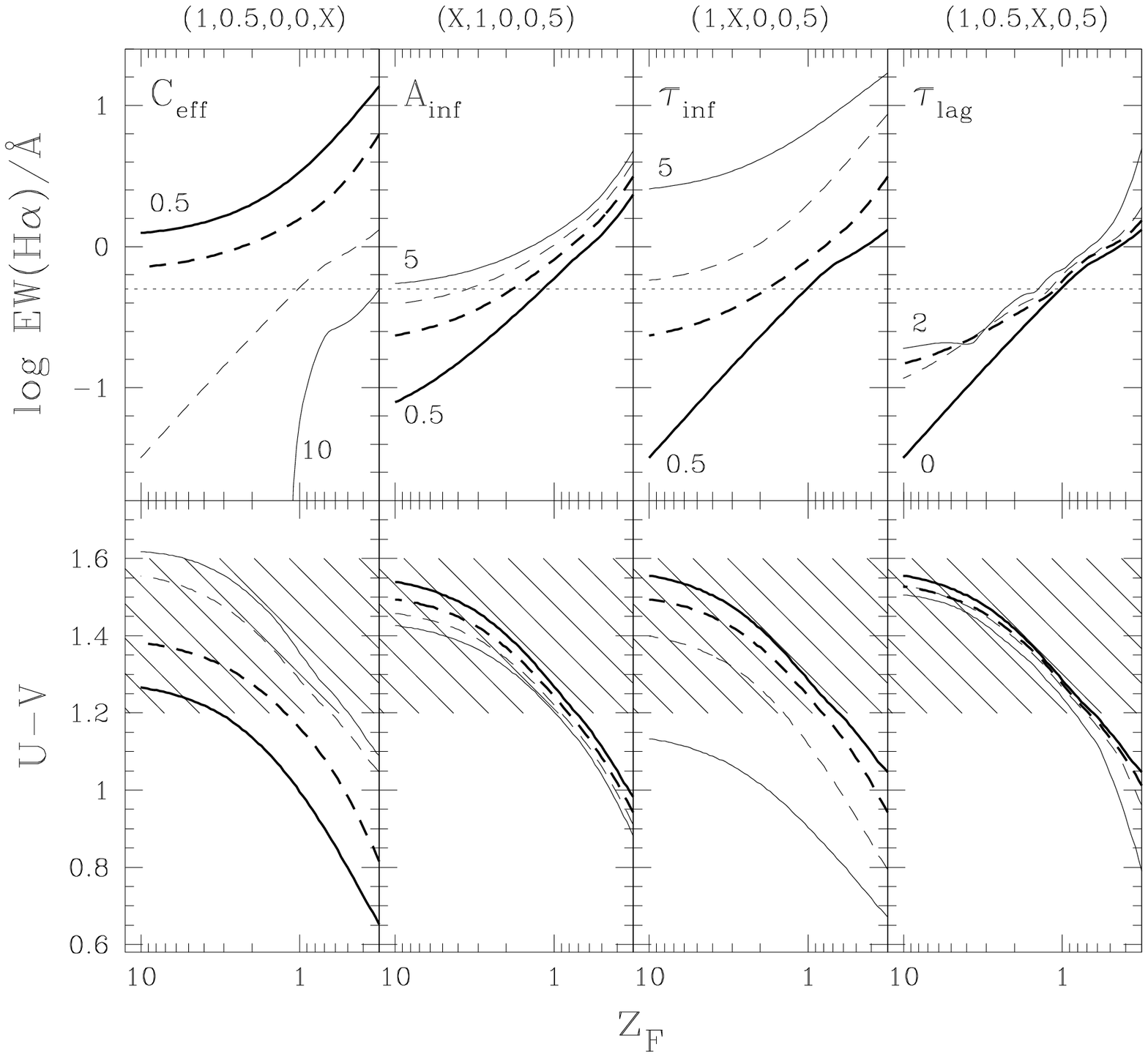}
\figcaption[fs_f2.eps]{{\sl Bottom Panels:} 
Projection of chemical enrichment tracks on rest frame $U-V$ color vs.
formation redshift.An open cosmology with $\Omega_0=0.3$ 
and $H_0=60$ km s$^{-1}$ Mpc$^{-1}$ is used to translate between ages 
and redshifts. The shaded region encompasses the color range found in 
local clusters such as Coma or Virgo. {\sl Top Panels:} Equivalent
widths of H$\alpha$ from the computed star formation rate (SFR) 
(see text for details). Some of the models can be ruled out as 
they would predict detectable emission lines in nearby early-type 
cluster galaxies. The dotted line represents an observational 
threshold at 0.5\AA . The leftmost panel shows the star formation 
efficiency ($C_{eff}$) must be high to avoid this. The numbers on top 
label the parameters used for each trajectory as 
$(A_{inf},\tau_{inf},\tau_{lag},B_{out},C_{eff})$ with $X$
meaning several values for that parameter are considered.\label{f2}}
\vskip+0.2truein

%%%%%%%%%%%%%%%%%%%%%%%%%%%%%%%%%%%%%%%%%%%%%%%%%%%%%%%%%%%%%

\noindent
Models that satisfy this constraint must have a high star formation
efficiency which implies a strong and short-lived burst of star 
formation, in agreement with the standard scenario.

\subsection{Analytic Approximation}
An analytic solution to a simplified version of the chemical
enrichment equations can illustrate the importance of outflows
in the final metallicities. Let us assume the Instantaneous
Recycling Approximation with stellar lifetimes 0 and $\infty$ separated
at some mass cutoff $M_0$. Let us also consider no infall. 
In this case, the integration of the gas mass equations (3) and (5)
is an exponential decay:
\begin{equation}
\mu_g(t)= e^{-t/T_g},
\end{equation}
where $T_g^{-1}=C_{eff}[1-R (1-B_{out})]$, and $R$ 
is the returned fraction of gas, 
i.e. $R = \int_{M_0}^\infty dm\phi (m)(m-w_m)$,
dependent on the IMF. The metallicity of the gas can be readily solved
from (8) and (10) giving:
\begin{equation}
Z_g(t)=1-e^{-\gamma t}\sim\gamma t,
\end{equation}
where $\gamma = C_{eff}(1-B_{out})(1-R)y$, and 
$(1-R)y = \int_{M_0}^\infty dm\phi(m)mp_m$, i.e. dependent both on the
IMF and the stellar yields. The asymptotic behavior 
($Z_g(t)\rightarrow 1$) is  formally correct but physically
meaningless, as the real equations should take into account the
change of the yields with metallicity. We have  assumed
$Z_g(t) \ll 1$ for the approximation. Equations (11) and (12) 
show that the final metallicities are of order:
\begin{equation}
\gamma T_g = {(1-B_{out})(1-R)y\over 1-(1-B_{out})R},
\end{equation}
and are  independent of the star formation efficiency. This gives some
sort of ``effective yield'' to which the metallicity will tend at
late times, which depends on the IMF (through $R$, and $y$),
the stellar yields ($y$) and the ejected fraction ($B_{out}$). These
gas metallicities can be computed for a Salpeter and a Scalo IMF
with a cutoff at $M_0=10M_\odot$, giving $2.4Z_\odot$ and 
$0.7Z_\odot$, respectively. A Salpeter IMF is flatter at the high
mass end, which explains the higher metallicity. Hence, a change
in the slope of the IMF can also reproduce the same metallicity range
found for a range of outflows.

\subsection{Caveats}
Before venturing on the predictions generated by this simple chemical
enrichment model, it is worth mentioning the most important
uncertainties and simplifications which might contribute to changing the
output. First and foremost, the large uncertainties in modelling
stellar evolution result in large error bars in the yields ($p_m$) 
which will render any estimation of chemical evolution only believable 
on a qualitative
basis regardless of the fine-tuning of the chemical enrichment prescriptions
that are 
considered. This is what motivates a simple recipe that takes into account 
the most basic mechanisms driving chemical enrichment in galaxies, as well
as tracing only the net metal content rather than following the abundance
of single elements. The connection between ejected fractions
and absolute luminosities should not be taken at face value but, 
rather, as a rough estimate. There are quite a few
factors affecting the asymptotic metallicities (as seen in the
previous section), such as changes in the IMF slope, the dependence 
of the stellar yields on metallicity (e.g. Portinari, Chiosi \& 
Bressan 1998\markcite{por98}) or the metallic composition of the
ejected winds if mixing is not perfect.

Finally, any non-primordial infall will add a 
degeneracy between infall parameters and ejected fractions: the infall
of gas with a significant metal abundance will mimic the same enrichment
tracks for primordial infall in galaxies with lower ejected fractions. 
However, if an anti-correlation between absolute luminosity and
outflows is considered, then we could argue that such a mechanism will 
make some of the fainter galaxies as red as the brightest ones, thereby
increasing the scatter about the color-magnitude relation. 

Therefore, we emphasize that this model should be considered as
a ``back-of-the-envelope'' numerical simulation, exploring an
important region of the enormous volume of parameter space in 
the process of galaxy formation and evolution. In principle, 
the most fundamental processes are the ones considered in this
paper, namely the role of outflows, infall of gas or a 
significant variation in the efficiency forming stars. 
The role of other mechanisms mentioned above and elsewhere
(e.g. pre-enrichment, top-heavy IMFs or variable stellar yields) 
will be left for later study.

%%%%%%%%%%%%%%%%%%%%%%%%%%%%%%%%%%%%%%%%%%%%%%%%%%%%%%%%%%%%%%%%%%%%%%%%%%%%%

\section{Simulating a Cluster}
Once the chemical enrichment tracks have been computed for a single galaxy,
the process can be extended to simulate the population of early-type
galaxies in a cluster. A suitable connection between the ejected 
fraction (parameter $B_{out}$) and the absolute luminosity is obtained by
using the CM relation of a local cluster as a constraint.
Hence, for a given value of the ejected fraction, the chemical enrichment
model is used to infer a metallicity for a given age. 
Figure~3 shows the relation between luminosity, stellar mass and
color vs. the ejected fraction as constrained by the color-magnitude 
relation of local clusters.
The range of galaxies observed in the Coma cluster by Bower, Lucey \& Ellis
(1992)\markcite{ble92}, which corresponds to absolute luminosities 
$-23.5<M_V<-19.5$ maps into a range of ejected fractions: $0<B_{out}<0.5$.
The dependence of $U-V$ an $V-K$ colors with $B_{out}$ is also plotted for a set
of five toy models described in table~1. The fiducial model ($A1$) assumes
no infall and a star formation efficiency of $C_{eff}=10$. The other models
include infall of gas. Model $C$ has a reduced star formation efficiency
and model $D$ delays the infall 1 Gyr after the galaxy starts forming stars
($\tau_{lag} = 1$ Gyr).
All these models are degenerate with respect to color, which can be explained
with the enrichment tracks shown in figure~1: only parameter $B_{out}$ yields
a significant range of metallicities. So, these different models will just
slightly ``modulate'' the enrichment tracks of the bottom left panel of 
figure~1.

\begin{table*}
   \begin{center}
   \caption{Chemical Enrichment Toy Models\label{tab1}}
   \vskip+0.5truecm	
   \begin{tabular}{c|ccccc|l}\hline\hline
Model & $A_{inf}$ & $\tau_{inf}$ & $\tau_{lag}$ & $B_{out}$ 
& $C_{eff}$ & Comments \\
  \hline
 $A1$ & 1 & 0.5  &  0  & $0-0.95$ & 10 & Default \cr
 $A2$ & 1 &  1   &  0  & $0-0.95$ & 10 & Extended Infall \cr
 $B$  & 0 & ---  & --- & $0-0.95$ & 10 & No Infall \cr
 $C$  & 1 &  1   &  0  & $0-0.95$ &  5 & Low SF Efficiency \cr
 $D$  & 1 & 0.5  &  1  & $0-0.95$ & 10 & Delayed Infall \cr
  \hline\hline
   \end{tabular}
   \end{center}
 \end{table*}

As long as high enough formation redshifts are considered ($z_F\simgt 3$), 
we can assume a pure metallicity sequence for the local cluster that is 
used as a constraint. This approximation is very robust since more realistic 
non-pure metallicity sequences will still give the same metallicity 
range: figure~1 shows
that after a few billion years the metallicity stays roughly constant:
once the most massive stars ($M\simgt 10M_\odot$) have reached the end
of their lives, chemical enrichment is controlled by the 
population of intermediate mass stars which have long lifetimes as
well as low metal yields. For a given ejected fraction, we can
read off the age and metallicity and --- using the latest popu-

%%%%%%%%%%%%%%%%%%%%%%%%%%%%%%%%%%%%%%%%%%%%%%%%%%%%%%%%%%%%%
%%%%%%%%%%%%%%%%%%%%%%%%   FIG. 3   %%%%%%%%%%%%%%%%%%%%%%%%%
%%%%%%%%%%%%%%%%%%%%%%%%%%%%%%%%%%%%%%%%%%%%%%%%%%%%%%%%%%%%%

\centerline{\null}
\vskip3.4truein
\includegraphics{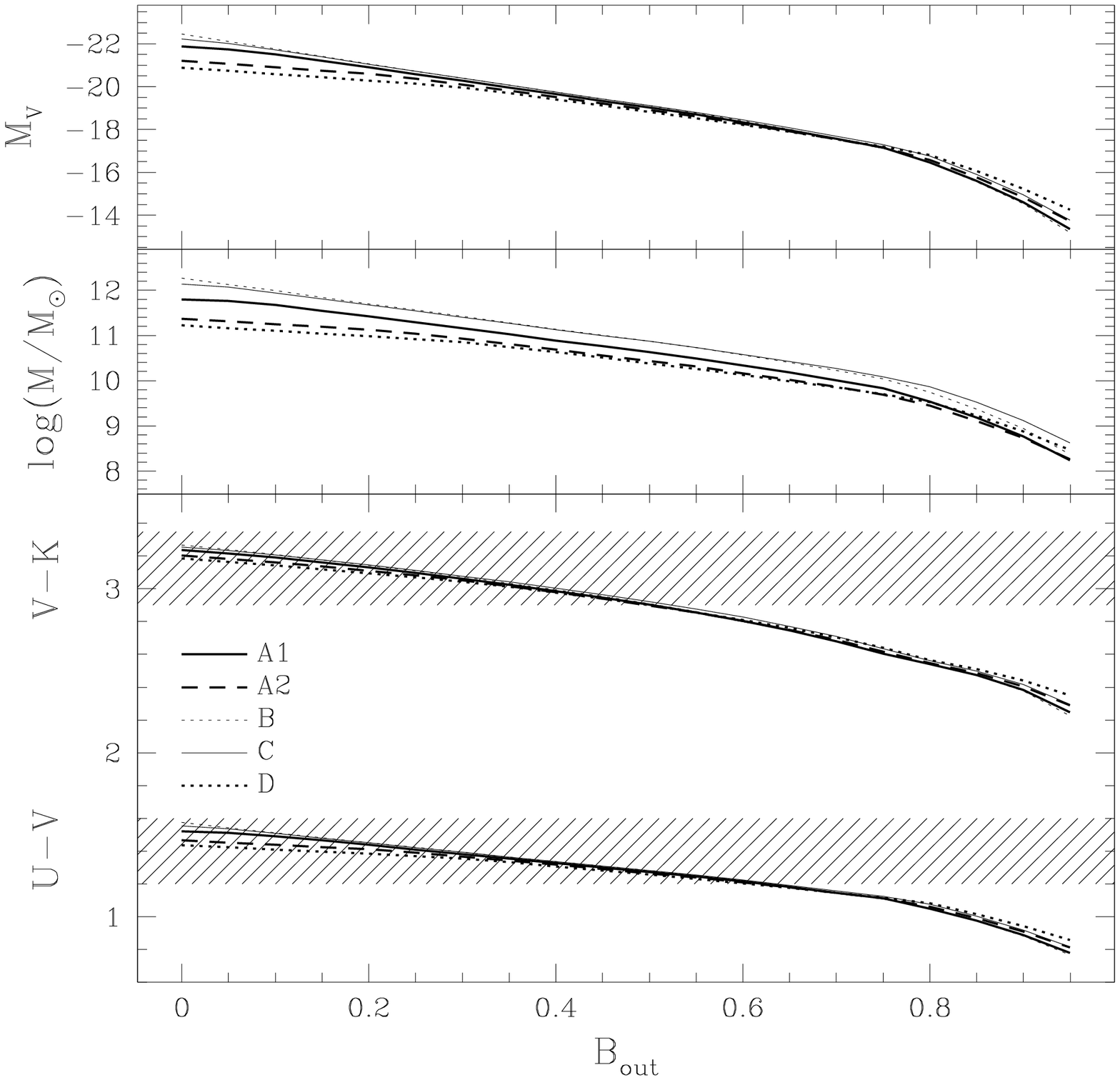}
\figcaption[fs_f3.eps]{Evolution of $V$-band absolute luminosity,
stellar mass, $U-V$ and $V-K$ color (both computed in the rest frame), 
with respect to the ejected fraction parameter --- $B_{out}$ --- for 
a set of enrichment toy models (see table~1). The $M_V$ evolution is just a
mapping of the $U-V$ color curves, since the Color-Magnitude
relation of the Coma cluster from Bower, Lucey \& Ellis (1992) 
is used as constraint. The curves for all five toy models are 
degenerate, so that only parameter $B_{out}$ is useful in order to identify
the early-type galaxies along the red envelope sequence.\label{f3}}
\vskip+0.2truein

%%%%%%%%%%%%%%%%%%%%%%%%%%%%%%%%%%%%%%%%%%%%%%%%%%%%%%%%%%%%%

\noindent
lation synthesis models from Bruzual \& Charlot (1999)
(hereafter BC99) --- the rest frame $U-V$ color. Finally, the
CM relation from BLE92 is considered in order to infer an absolute 
luminosity. The sample of galaxies is chosen to fit the
the luminosity function of early-type (normal plus dwarf) galaxies in
the Virgo cluster measured by Sandage, Binggeli \& Tammann (1985)
\markcite{sa85}, using a Schechter function with parameters $\alpha = -1.40$
and $M_V = -23.0$ (a translation from $B$ to $V$ band was performed
assuming a color $B-V\sim 1.0$ for an early-type galaxy).

%%%%%%%%%%%%%%%%%%%%%%%%%%%%%%%%%%%%%%%%%%%%%%%%%%%%%%%%%%%%%%%%%%%%%%%%%%%%%

\section{Color-Magnitude Relation}
One of the most useful estimators of evolution in cluster early-type
galaxies is the color-magnitude relation. The study of rest frame
$U-V$ color versus absolute luminosity traces the change in the age
and metallicity of the stellar populations. The small scatter found
in clusters at low and moderate redshifts is an indicative sign of either
a highly synchronous formation process or a very stable region 
in color-magnitude space. 

Figure~4 shows the prediction of the CM relation at four different
redshifts for the three star formation scenarios described in table~2:
The M-model is a monolithic collapse process 
(e.g. Larson \& Tinsley 1974\markcite{lt74}), where the formation 
redshift of all early-type galaxies follows a Gaussian distribution
with mean $z_F=5$ and standard deviation $\sigma (z_F) = 1$, thus
written as M(5,1). The H-model represents a cluster in which most of 
the star formation in the brightest galaxies occur at late times
compared to the faintest ones. This model cannot be directly related to
a specific hierarchical clustering scenario 
(e.g. Kauffmann, White \& Guiderdoni 1993\markcite{kwg93}), 
although it resembles a scenario where strong
mergers take place in the brightest galaxies at late times, with a
large fraction of stars being formed at lower redshifts, assumed to be
$z_F\simgt 2$, whereas the faintest galaxies have formation
redshifts $z_F\sim 10$. The correlation
between luminosity and formation redshift is assumed to be linear and
with a gaussian spread of $\sigma = 1$. The notation for this model is
H(2,10,1). We do not assume a major stage of
star formation later than $z_F\sim 2$ as current observations of 
early-type cluster galaxies in a wide redshift range ($0<z<1$) 
seem to discard that possibility 
(e.g. Stanford et al. 1998\markcite{sed98}). 

Finally, the IH-model is an inverted
hierarchical clustering process which corresponds to H(10,2,1), i.e. the
brightest galaxies are the first ones to form at $z_F\sim 10$. 
This type of evolution will occur if the first objects to form
were to reheat the intergalactic medium, raising the Jeans mass
so that the more massive galaxies form first (Blanchard, 
Valls-Gabaud \& Mamon 1992\markcite{bla92}).
Table~2 also lists the time lapse over which star formation
takes place in the cluster. A monolithic collapse model condenses this
stage into 0.6 Gyr, whereas the H and IH models 

%%%%%%%%%%%%%%%%%%%%%%%%%%%%%%%%%%%%%%%%%%%%%%%%%%%%%%%%%%%%%
%%%%%%%%%%%%%%%%%%%%%%%%   FIG. 4   %%%%%%%%%%%%%%%%%%%%%%%%%
%%%%%%%%%%%%%%%%%%%%%%%%%%%%%%%%%%%%%%%%%%%%%%%%%%%%%%%%%%%%%

\centerline{\null}
\vskip3.4truein
\includegraphics{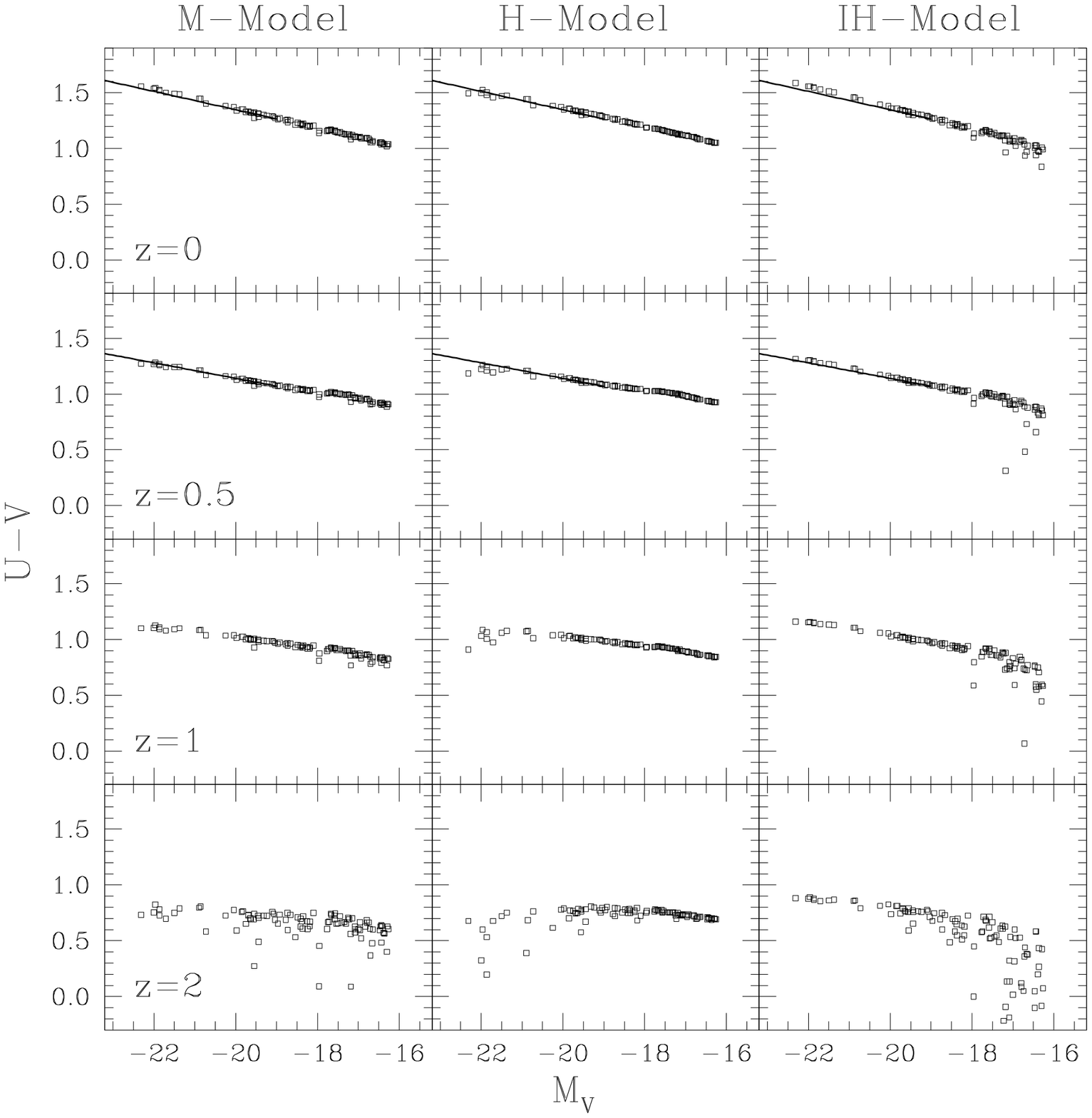}
\figcaption[fs_f4.eps]{Simulation of the rest frame $U-V$ vs $M_V$ 
color-magnitude relation of a cluster at different redshifts for 
three star formation models: Monolithic ($M$), Hierarchical ($H$) 
and Inverted Hierarchy ($IH$) (see table~2).
The solid lines are linear fits to observed CM relations at
$z\sim 0$ (Coma, Bower, Lucey \& Ellis 1992) and $z\sim 0.5$
(Cl0016+16, Ellis et al. 1997). Notice the presence of blue 
outliers with respect to absolute luminosity at high redshift 
depends on the formation scenario: A ``bottom-up'' hierarchy could 
thus be tested by the observation of the bright end of the
color magnitude relation in high redshift clusters.\label{f4}}
\vskip+0.2truein

\begin{table*}
   \begin{center}
   \caption{Star Formation Scenarios\label{tab2}}
   \vskip+0.5truecm
   \begin{tabular}{c|cc|l}\hline\hline
Label & Model & $\Delta t_{\rm SF}$/Gyr & Comments \\
      &       & ($\Omega_0=0.3 ; h_0=0.6$) & \\
  \hline
 $M$  & M(5,1)    & 0.60 & Monolithic\cr
 $H$  & H(2,10,1) & 5.01 & Hierarchical\cr
 $IH$ & H(10,2,1) & 5.01 & Inverted Hierarchy\cr
  \hline\hline
   \end{tabular}
   \end{center}
 \end{table*}

%%%%%%%%%%%%%%%%%%%%%%%%%%%%%%%%%%%%%%%%%%%%%%%%%%%%%%%%%%%%%

\noindent
extend the interval to
5 Gyr. Taking into account the luminosity function, the inverted hierarchy 
model implies a rather strong rate of star formation at low redshift.
The difference between these three star formation scenarios can be seen 
in the CM predictions at four redshifts in figure~4. The solid lines
are the linear fits to observed CM relations at $z\sim 0$ (Coma, BLE92),
and at $z\sim 0.5$ (Cl0016+16, Ellis et al. 1997\markcite{el97}).
One can see that 
these models are degenerate in clusters at low to moderate redshifts
($0<z<1$), taking into account observational uncertainties. However,
both hierarchical scenarios will yield a
significant population of blue outlier galaxies at redshifts $z>1$. 
The monolithic model produces a ``luminosity-blind'' scatter 
that increases at high redshift blueward of the red envelope all along 
the luminosity sequence, whereas the H and IH models preferentially 
display outliers at the bright and faint ends, respectively of the 
``main sequence'' red envelope. Thus, observations of high redshift 
clusters should  easily confirm or rule out an ordered hierarchy 
--- as opposed to a monolithic collapse --- in the formation of the 
stellar populations with respect to the total luminosity of the galaxy.
Notwithstanding possible dynamical processes that might have triggered 
the observed recent star formation, we can identify these blue 
galaxies as young spheroids with a significant population
of A-type stars (there should be no massive OB stars as long as we assume 
a negligible star formation rate at $z<2$), so that they will plausibly 
display the spectral feature of a post-starburst galaxy (also referred to 
as E+A or k+a) (Dressler \& Gunn  1983\markcite{dg83}), 
i.e. a spectrum of an old population of stars (E or k) with an added 
population of young stars (A) which will be characterized by strong 
Balmer absorption lines. This issue will be treated in more detail 
in \S7 as this population can represent an important tracer of 
cluster evolution.

The main parameters of the CM relation (slope, zero point measured 
at the brightest cluster galaxy and the  rms residuals about the linear fit)
can be used to constrain the epoch of star formation in galaxies as
well as the basic factors contributing to their evolution. 
Figures~5a and 5b show the evolution of the main CM parameters as a 
function of redshift both for several enrichment toy models (fig~5a)
and for three star formation models (fig~5b). Figure~5a
shows that a wide range of infall rates, timescales or infall delays
as well as a stronger star formation efficiency result in no
major change from the default model: an extension of the result 
shown for the evolution of rest frame $U-V$ color for a single
galaxy with different chemical enrichment processes (fig~2). Only a large
difference in ages will yield different slopes, zero points or
residuals. However, the need for strong initial star formation
rates imply a negligible difference between models unless 
redshifts very close to $z_F$ are considered. Only an IH model
yields significantly different residuals about the CM relation
as well as a steeper slope at high redshift.

%%%%%%%%%%%%%%%%%%%%%%%%%%%%%%%%%%%%%%%%%%%%%%%%%%%%%%%%%%%%%
%%%%%%%%%%%%%%%%%%%%%%%%   FIG. 5   %%%%%%%%%%%%%%%%%%%%%%%%%
%%%%%%%%%%%%%%%%%%%%%%%%%%%%%%%%%%%%%%%%%%%%%%%%%%%%%%%%%%%%%

\begin{figure*}
\figurenum{5}
\plotfiddle{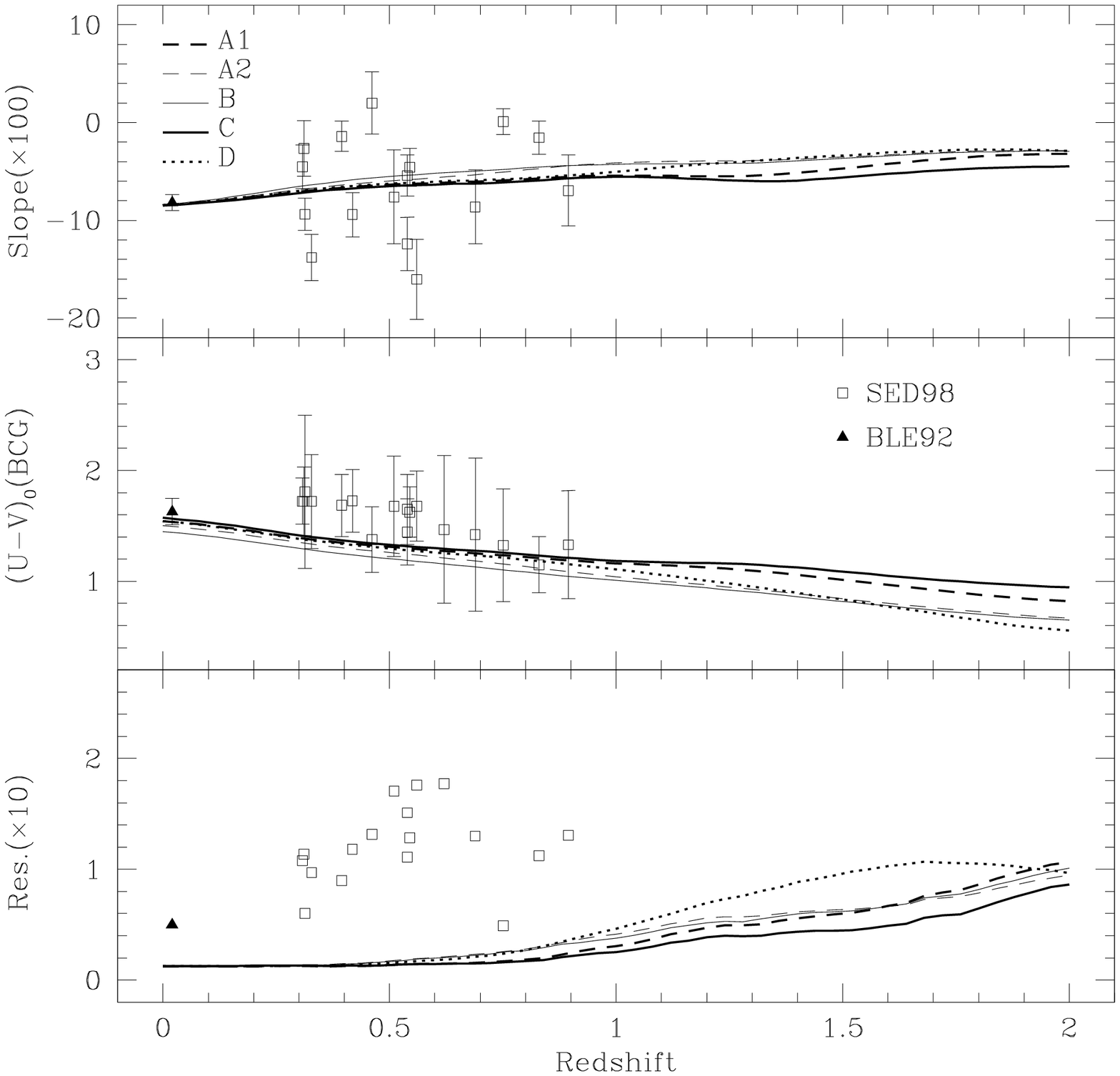}{425pt}{0}{40}{40}{-270}{150}
\plotfiddle{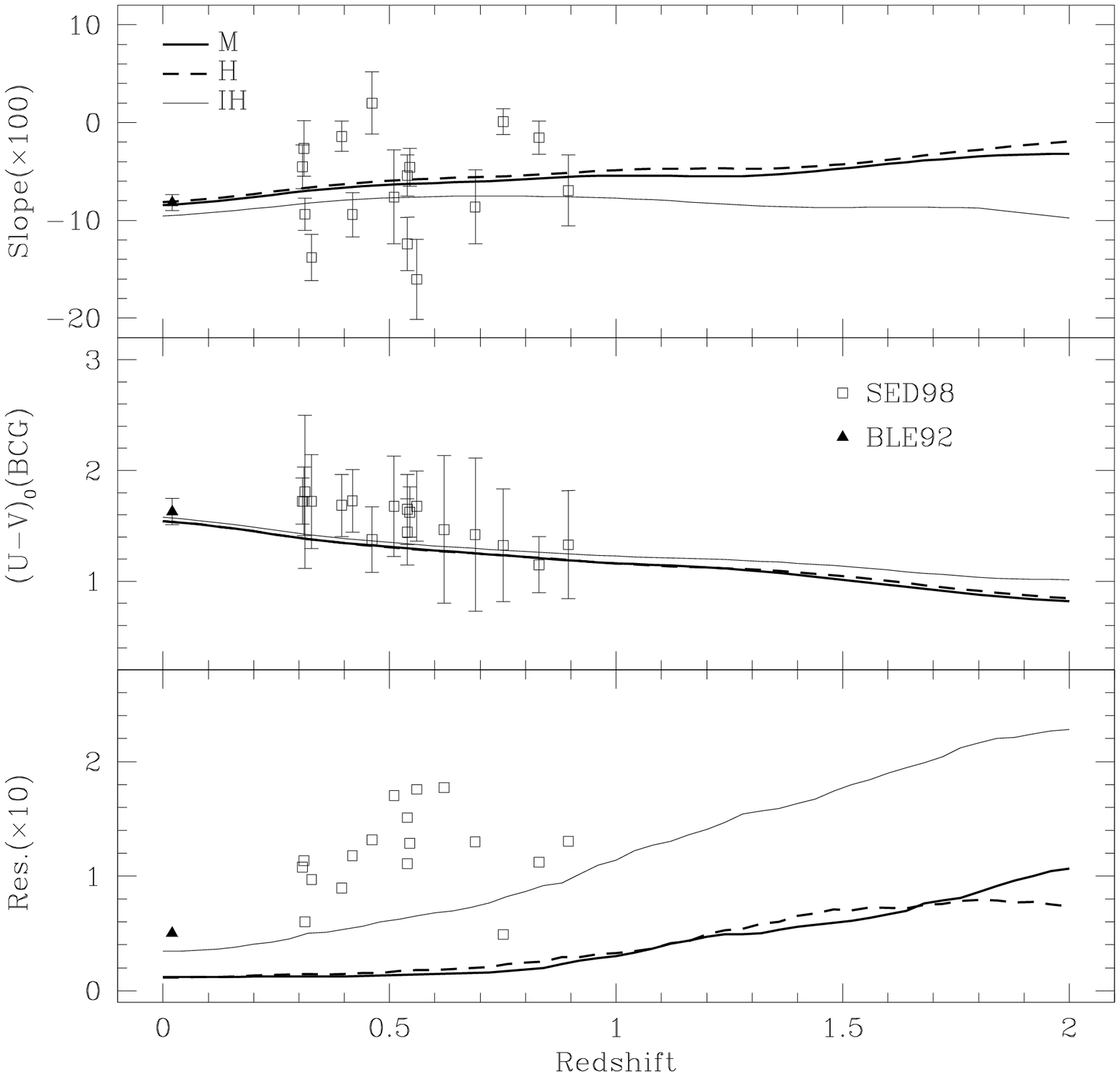}{425pt}{0}{40}{40}{0}{590}
\vspace{-22cm}
\vskip-0.5truein
\caption{{\it a)} Evolution of the main parameters in the
Color-Magnitude relation (slope, zero point evaluated at the
brightest cluster galaxy and rms residuals) for a set of enrichment
toy models. The points are cluster observations from Bower, Lucey 
\& Ellis (1992) (triangle), and Stanford, Eisenhardt \& Dickinson 
(1998) (squares). For SED98 we translated their colors 
(which used different filters depending on the cluster redshift
to trace the 4000\AA\  break) to actual rest frame $U-V$, using
passive evolution with a formation redshift $z_F=5$ along with
the models from Bruzual \& Charlot (In preparation). The rms 
residuals from the observations should be taken with caution since 
cluster membership is not confirmed for all early-type galaxies 
observed in SED98.{\it b)} Same as figure~5a with respect to three
different star formation scenarios: Monolithic (M), Hierarchical
Clustering (H) and an Inverted Hierarchy (IH) 
(see table~2).\label{f5}}
\end{figure*}

Also shown in these two figures are the observations of 17 
clusters in the range $0.3<z<0.9$ from the sample of Stanford et al. 
(1998)\markcite{sed98}, corrected to rest frame ``$U-V$''
versus absolute luminosity in the $V$-band. The translation was done 
by assuming passive evolution and a formation
redshift $z_F=5$ for all of the galaxies in each cluster. Notice the
large scatter and error bars which render the estimation of star
formation histories a rather qualitative quest until a large
enough sample of cluster observations become available. The rms
residuals from the observed sample can only yield weak limits on the
intrinsic residuals from age and metallicity differences between
galaxies. Photometric uncertainties overwhelm the instrinsic
scatter unless high redshift clusters with a significant population
of young galaxies are observed. Furthermore, the sample of galaxies
has only been culled with regard to morphology by using high resolution
{\sl HST/WFPC2} images, but no spectroscopic confirmation of cluster
membership has been done for all the galaxies. Hence, a significant
contamination from line-of-sight early-type galaxies can be expected.
We should emphasize that the model residuals computed are a poor 
estimator of a hierarchical clustering process. In this scenario, 
the young outlier galaxies are the few brightest ones, which do not 
count as much towards the final residuals as the more numerous older 
and fainter population. Hence, luminosity-weighted residuals would 
be more useful for this purpose.

%%%%%%%%%%%%%%%%%%%%%%%%%%%%%%%%%%%%%%%%%%%%%%%%%%%%%%%%%%%%%%%%%%%%%%%%%%%%%

\section{Mass-to-Light ratios}
The stellar mass-to-light ratios can also be computed using the population
synthesis models from BC99. The ratio of mass to luminosity is a very
age-sensitive quantity, thereby making it one of the most useful
observables for breaking the age and metallicity degeneracy.
Unfortunately, observed $M/L$ ratios have large uncertainties so that
a large sample in a wide range of redshifts is necessary in order to
discriminate between formation scenarios. Figures~6a and 6b show the
predictions from our chemical enrichment model for the main parameters
of the correlation between $\log M/L_V$ and $\log M$.
A non-zero slope yields the tilt of the fundamental plane relative to
the expectation using the virial theorem. A range of chemical enrichment
toy models results in a degeneracy for the slope, zero point and
residuals of this correlation. In analogy with the CM relation, these
parameters are most sensitive to age differences as shown in figure~6b
for three star formation scenarios. The middle panels show there is good 
agreement in the zero points with $M/L$ observations measured for five
clusters: Coma ($z=0.02$), Cl1358+62 ($z=0.33$), Cl0024+16 ($z=0.39$), 
MS2053+03 ($z=0.58$) and MS1054-03 ($z=0.83$), obtained from the 
compilation of Van Dokkum et al. (1998)\markcite{vdk98}. The trend
with redshift gives a linear fit $\Delta\log M/L_V=-0.34z$ for the range
$0<z<1$, in agreement with Van Dokkum et al. (1998)\markcite{vdk98}, who
find $\Delta\log M/L_B=-0.4z$. However, we can see in the figures that
this evolution is not linear, and so a fit using a wider range of 
redshifts $0<z<2$ gives a flatter slope $\Delta\log M/L_V=-0.28z$. The
residuals could represent a way of discerning the formation history, but
given realistic values for the error bars, we doubt that cluster observations,
at least in the redshift range $0<z<1$ could be used for this purpose. 

A comparison of the slopes obtained with purely stellar masses
--- around $M/L_V\propto M^{0.075-0.085}$ for a monolithic model --- 
with actual observations: $M/L\propto M^{0.15-0.25}$ 
(e.g. Mobasher et al. 1999\markcite{mob99}) shows that stellar masses alone 
cannot account for this, and dark matter (DM) should therefore
be included. The slope
mismatch is caused by a correlation between stellar mass and dark matter,
so that there is more invisible mass in galaxies with more mass in stars.
Given that the model presented in this paper only gives the average,
qualitative trend of different observables, we should only take these slopes
as indicative. However, if we speculate that the actual mismatch between
mass in stars and DM is similar to the one obtained here (i.e. between a 
slope of $\sim 0.15-0.25$ for DM and $\sim 0.075-0.085$ for stars), then the
trend should be $M_{\rm DM}\propto M_{\rm St}^\alpha$, with
$\alpha=1.15\pm 0.08$, i.e. the scaling between dark matter and luminous
matter is roughly linear, which implies a mass-independent bias for
ellipticals. This renders the studies of (stellar) $M/L$ ratios using 
population synthesis models applicable to dynamical masses as well, with the
proviso that an offset should be added when dealing with the total mass.

%%%%%%%%%%%%%%%%%%%%%%%%%%%%%%%%%%%%%%%%%%%%%%%%%%%%%%%%%%%%%
%%%%%%%%%%%%%%%%%%%%%%%%   FIG. 6   %%%%%%%%%%%%%%%%%%%%%%%%%
%%%%%%%%%%%%%%%%%%%%%%%%%%%%%%%%%%%%%%%%%%%%%%%%%%%%%%%%%%%%%

\begin{figure*}
\figurenum{6}
\plotfiddle{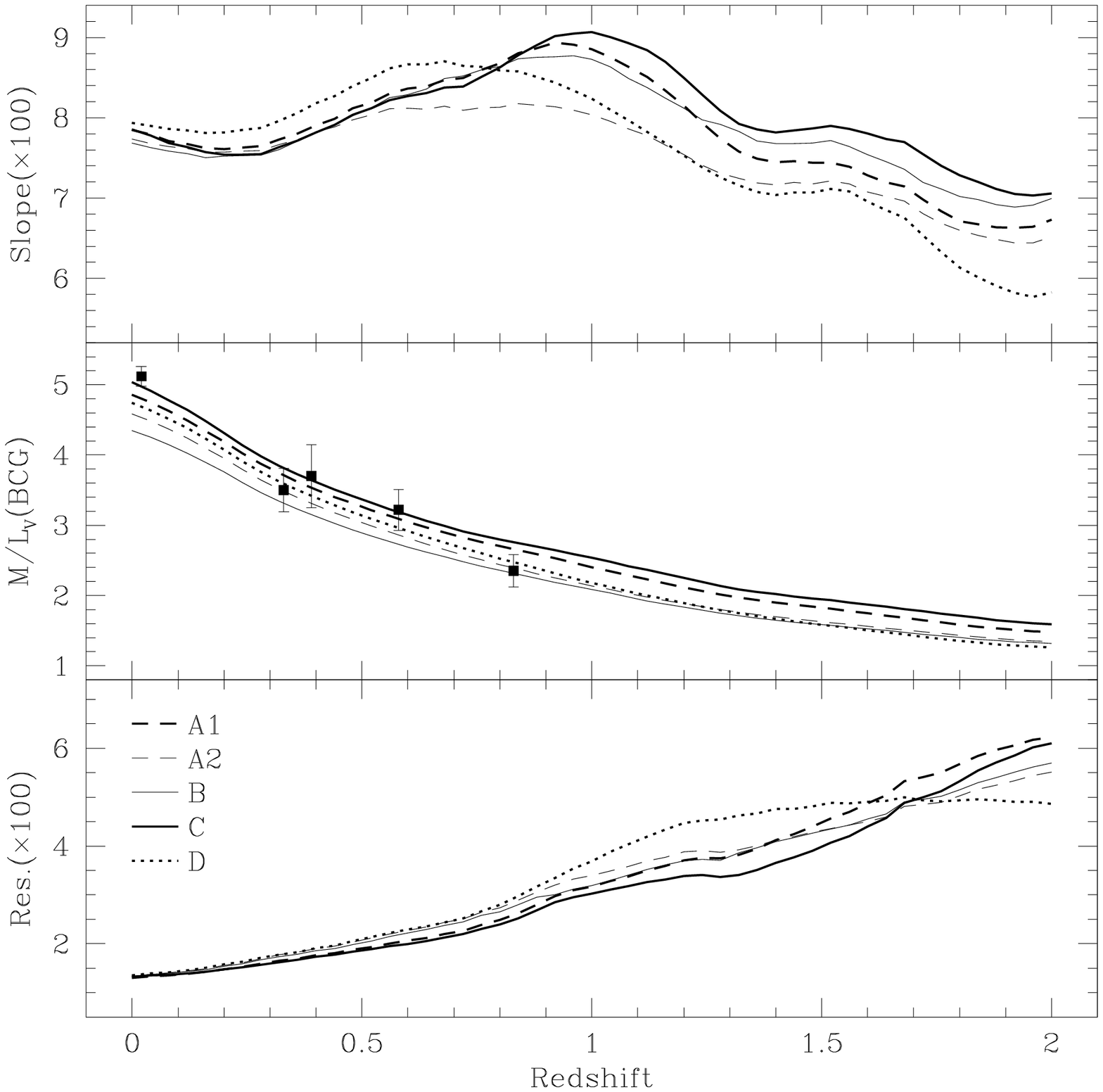}{425pt}{0}{40}{40}{-270}{150}
\plotfiddle{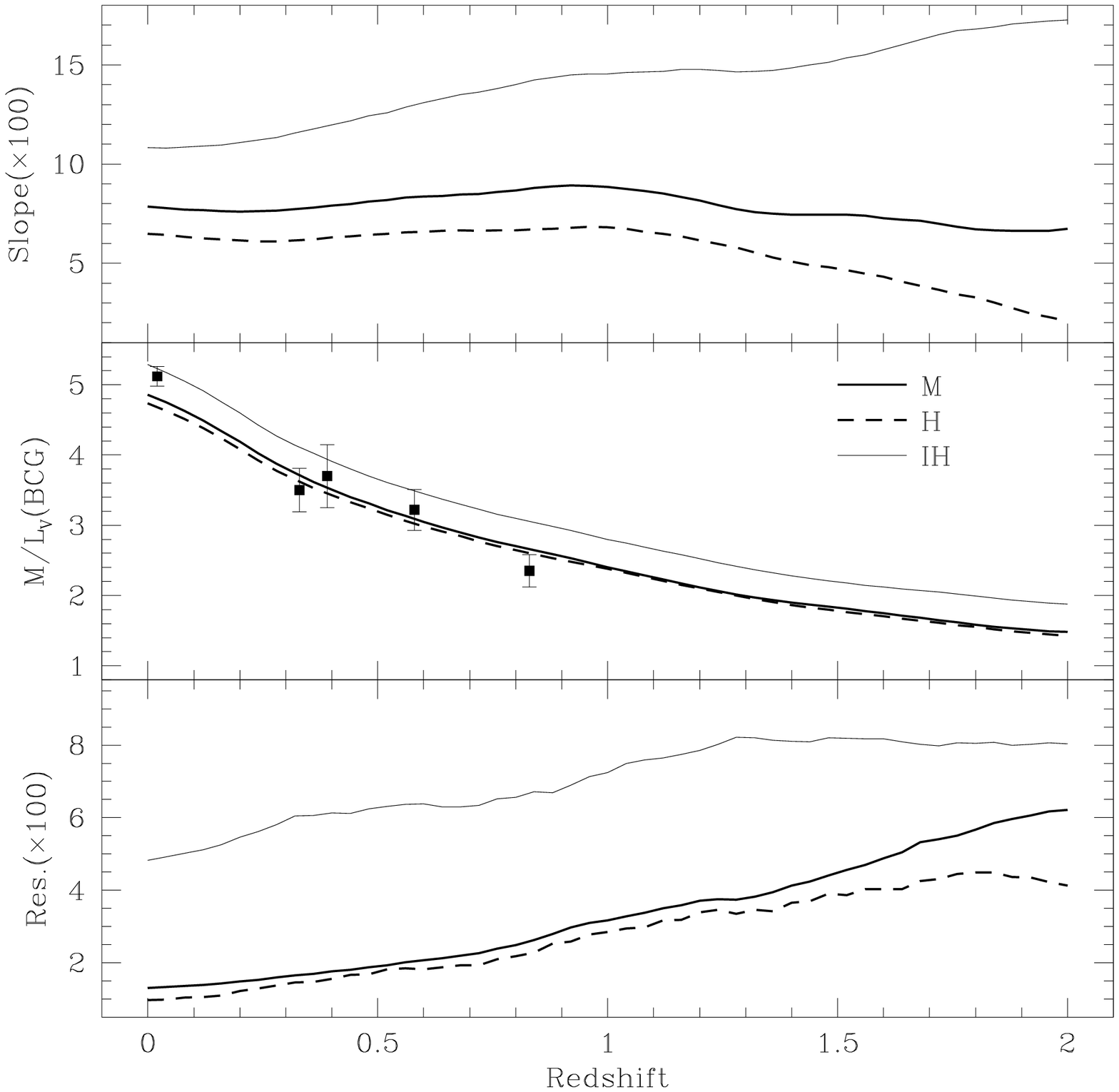}{425pt}{0}{40}{40}{0}{590}
\vspace{-22cm}
\vskip-0.5truein
\caption{{\it a)} Evolution of the slope, zero point at the
brightest cluster galaxy and residuals of the $\log (M/L_V)$ vs $\log M$
relation for a set of enrichment toy models. The points are values
for several clusters: Coma ($z=0.02$); Cl 1358+62 ($z=0.33$);
Cl 0024+16 ($z=0.39$); MS 2053+03 ($z=0.58$) and MS 1054-03 ($z=0.83$)
 from the compilation of figure~3 from Van Dokkum et al. (1998)
The slope --- around 0.075--0.085 --- for stellar mass-to-ligh ratios 
disagrees with actual observations of the TOTAL mass-to-light ratio
(0.15--0.25, e.g. Mobasher et al. 1999) which implies a correlation 
between total and stellar mass of order: 
$M_{\rm TOT}\propto M_{\rm Stellar}^{1.15\pm 0.08}$
{\it b)} Same as figure~6a with respect to three different
star formation scenarios (see table~2).\label{f6}}
\end{figure*}

%%%%%%%%%%%%%%%%%%%%%%%%%%%%%%%%%%%%%%%%%%%%%%%%%%%%%%%%%%%%%%%%%%%%%%%%%%%%%

\section{The Evolution of the Magnesium Abundance}
The abundance of magnesium traces the star formation history of the
population of massive stars in the galaxy. It is strongly correlated with
the central velocity dispersion ($\sigma$) which implies an enhanced star 
formation rate in more massive ellipticals. This correlation might possibly 
depend on the environment as found by Guzm\'an et al. (1992)\markcite{gu92} 
and J\o rgensen, Franx \& Kj\ae gaard (1996)\markcite{jo96}. Furthermore, 
there is an overabundance of magnesium in the cores of the brightest 
ellipticals which may imply an enhanced massive star formation rate 
in these regions. A solution to this overabundance would require 
an {\sl ad hoc} correction 
factor as a function of absolute luminosity. In order not to 
introduce further uncertainties,
we will simply show the predictions without any correction. Figure~7
shows the evolution with redshift of the Lick/IDS index $Mg_2$ 
(Worthey 1994)\markcite{wo94} versus the logarithm of the 
central velocity dispersion ($\log\sigma$)
for all three star formation scenarios
considered in this paper. The data points come from observations of local 
clusters by Guzm\' an et al. (1992, G92)\markcite{gu92}; 
J\o rgensen et al. (1996, J96)\markcite{jo96}; 
Pahre, Djorgovski \& de~Carvalho (1998, P98)\markcite{pah98}; 
and Colless et al. (1999, EFAR)\markcite{co99}, as well as from a few
elliptical galaxies in clusters at $z\sim 0.37$ observed by Ziegler \& Bender 
(1997, Z97)\markcite{zi97}. In this case, a slope cannot be
obtained because of the small observational sample. Thus, the error bars 
instead give the range of observed magnesium abundances.
The overabundance problem is readily seen in
the zero point estimated at the brightest cluster galaxy, as well as
in the predicted flatter slopes. A different age spread in the stellar
populations among galaxies --- expected for H and IH models at high redshift
($z\simgt 1$) --- will translate into a large scatter, explained
by the fact that $Mg_2$ is a strongly metallicity-dependent observable.
Hence, at late times (when the metallicity is roughly constant with age)
the magnesium abundance will be controlled by fixed parameters such as
the stellar yields or the slope and cutoff of the upper mass segment of
the IMF. Only the fraction of outflows will contribute to explain a
range of abundances as a function of absolute luminosity (or
velocity dispersion, via the Faber-Jackson relation). It is only at early
stages --- when the metallicity changes significantly with age --- that we
expect to find a noticeable difference among star formation scenarios.

%%%%%%%%%%%%%%%%%%%%%%%%%%%%%%%%%%%%%%%%%%%%%%%%%%%%%%%%%%%%%%%%%%%%%%%%%%%%%

\section{Blue Outliers as tracers of cluster evolution}
Figure~4 shows that the presence of galaxies falling conspicuously blueward
of the ``main sequence'' red envelope represents an important estimator for
evolution in early-type cluster galaxies. This blueness is caused by the 
presence of young A-type stars associated with a recent star formation process.
The main sequence lifetime of these stars translates into a time lapse 
of 1 to 4~Gyr after which the last important star formation process
ceased. This interval is correlated with the equivalent width of
Balmer absorption lines. The lack of star formation is confirmed by the
absence of emission lines from hot OB stars. Hence, the spectrum of these 
blue outlier galaxies would thus be classified as ``E+A'' (Dressler \& Gunn 
1983\markcite{dg83}). Some of the brightest blue outlier galaxies have
been confirmed to be cluster members, with an E+A spectral type. Table~3
shows the three brightest early-type outliers in cluster Cl0016+16 ($z=0.545$) 
from the sample of Ellis et al. (1997)\markcite{el97} (the MORPHS 
collaboration), who performed the morphological classification with high
resolution images from {\sl HST/WFPC2}. All three galaxies have been 
confirmed to be cluster members as well as featuring an E+A spectral type, 
obtained either through spectroscopy (Dressler et al. 1999\markcite{dr99}; 
referred to in the table as MORPHS) or narrowband imaging (Belloni 
\& R\"oser 1996\markcite{br96}). The small solid angle covered by
{\sl WFPC2} makes meaningless any statistical estimation of field 
contaminants from morphologically segregated luminosity functions 
(e.g. Driver et al. 1998\markcite{dr98}).

%%%%%%%%%%%%%%%%%%%%%%%%%%%%%%%%%%%%%%%%%%%%%%%%%%%%%%%%%%%%%%%
%%%%%%%%%%%%%  Blue Outlier Members  %%%%%%%%%%%%%%%%%%%%%%%%%%
%%%%%%%%%%%%%%%%%%%%%%%%%%%%%%%%%%%%%%%%%%%%%%%%%%%%%%%%%%%%%%%
 \begin{table*}
   \begin{center}
   \caption{Blue Outlier Cluster Members in Cl0016+16 ($z=0.545$)\label{tab3}}
   \vskip+0.5truecm
   \begin{tabular}{c|cc|cccc|c}\hline\hline
ID       & R.A. & Dec. & F814W & F555W-F814W & EW($H\delta$) & $z$ & Source\\
(Morphs) & (J2000) & (J2000) & &           &  (\AA )      &     &  \\
  \hline
 266 & 00:18:32.22 & 16:25:06.6 & 20.35 & 2.04 & 8.5 & 0.5447 & MORPHS\\
2026 & 00:18:35.25 & 16:25:45.1 & 20.64 & 2.19 & --- & 0.53   & B\&R96\\
 809 & 00:18:31.22 & 16:26:52.1 & 20.67 & 2.06 & 5.3 & 0.5304 & MORPHS\\ 
  \hline\hline
   \end{tabular}
   \end{center}
 \end{table*}

The comparison of the measured equivalent widths for $H\delta$ 
with population synthesis models such as BC99 give an age estimate 
between 0.5 and 2 Gyr for a reasonable range of metallicities 
$0.2 < Z/Z_\odot < 2.5$, with the younger estimates corresponding 
to the higher metallicities. These estimates are obtained for a model
with a single burst. Models with multiple peaks in the star
formation rate will yield younger ages for the last major bursting
episode. Hence, the ages given above can be taken as upper limits.
The population of blue early-type outliers can thus
be considered as an important estimator of
cluster evolution. A comprehensive study could yield valuable 
information about the process of galaxy formation
in clusters. For instance, if we use the list of early-type galaxies in 
cluster Cl0016+16 observed by the MORPHS collaboration, and we assume
that all spheroids in the red envelope are cluster members, then 
we can state that down to a completeness magnitude of $F814W < 21.0$,
three out of 25 galaxies are blue outliers, with ages $t<2$ Gyr, that is
with a formation redshift $z_F\simlt 1$. If we assume a monolithic formation
process with a gaussian distribution and a mean at $z_F=3$, then this 
12\% fraction will imply a standard deviation of order $\sigma (z_F)\sim 2$,
i.e. a rather extended process of star formation. A hierarchical model,
where the stellar populations of the faintest galaxies are assembled first,
would thus be preferred. This simple calculation shows the 
power of ``outlier statistics'': measurements down to deep completeness
magnitudes in a sample of clusters comprising a wide range of
redshifts will pose severe constraints on the epochs of star formation.
However, we should emphasize that non-spheroidal morphologies might
also contribute to the fraction of galaxies that evolve into red
envelope early-types. Hence, the simple calculation presented here
would only represent a lower limit to the fraction of evolving 
early-type galaxies, comprising only those that appear already as
spheroidal.

%%%%%%%%%%%%%%%%%%%%%%%%%%%%%%%%%%%%%%%%%%%%%%%%%%%%%%%%%%%%%%%%%%%%%%%%%%%%%

\section{The Enrichment of the Intracluster Medium}
The intracluster medium (ICM) is a metal-rich hot gas which pervades the
central parts of galaxy clusters, and can be detected in X rays via thermal
bremsstrahlung radiation. The temperature of this gas is in the range
$2<kT_X<14$ keV and exceeds the mass of stars in galaxies by a factor of
2 to 10. The metallicity of the ICM is rather high, around $Z_\odot /3$,
showing no evidence of evolution up to redshifts $z\sim 0.5$
(Mushotzky \& Loewenstein 1997\markcite{ml97}). There is a strong 
correlation between the iron mass content of the ICM and the optical
luminosity (Arnaud et al. 1992\markcite{ar92}), a clear sign of a
connection between the metallicity of the ICM and the evolution of the
early-type population in the cluster. The model described in this paper
can be used in order to follow the chemical evolution of the gas and metals being 
ejected into the ICM. Given the connection found between absolute luminosity
--- or central velocity dispersion, using the Faber-Jackson relation 
(Faber \& Jackson 1976\markcite{fj76}) --- and the ejected fraction, we
can see that a low mass galaxy with $\sigma = 45$ km s$^{-1}$ or
absolute luminosity $M_V = -16.5$, will eject 90\% of its content, being 
practically destroyed right after the first stage of supernova explosions
by its massive constituent stars. Therefore, a model for studying the enrichment
of the ICM should take into account the contribution from the low mass
population of galaxies in the cluster (Dekel \& Silk 1986\markcite{dk86}).
Note that Mac~Low and Ferrara (1999)\markcite{mf99} argue that in thin 
disk galaxies, winds are effectively suppressed except in very low mass 
dwarfs; however for the early-type systems we are considering, the 
spheroidal geometry should render their argument ineffective.

Other mechanisms could also contribute significantly
to the metallicity of the ICM. For instance, if we were to consider the 
merging of disk galaxies with similar masses as precursors of the
brightest ellipticals (Toomre \& Toomre 1972\markcite{tt72}), then a large

%%%%%%%%%%%%%%%%%%%%%%%%%%%%%%%%%%%%%%%%%%%%%%%%%%%%%%%%%%%%%
%%%%%%%%%%%%%%%%%%%%%%%%   FIG. 7   %%%%%%%%%%%%%%%%%%%%%%%%%
%%%%%%%%%%%%%%%%%%%%%%%%%%%%%%%%%%%%%%%%%%%%%%%%%%%%%%%%%%%%%

\setcounter{figure}{6}
\centerline{\null}
\vskip3.4truein
\includegraphics{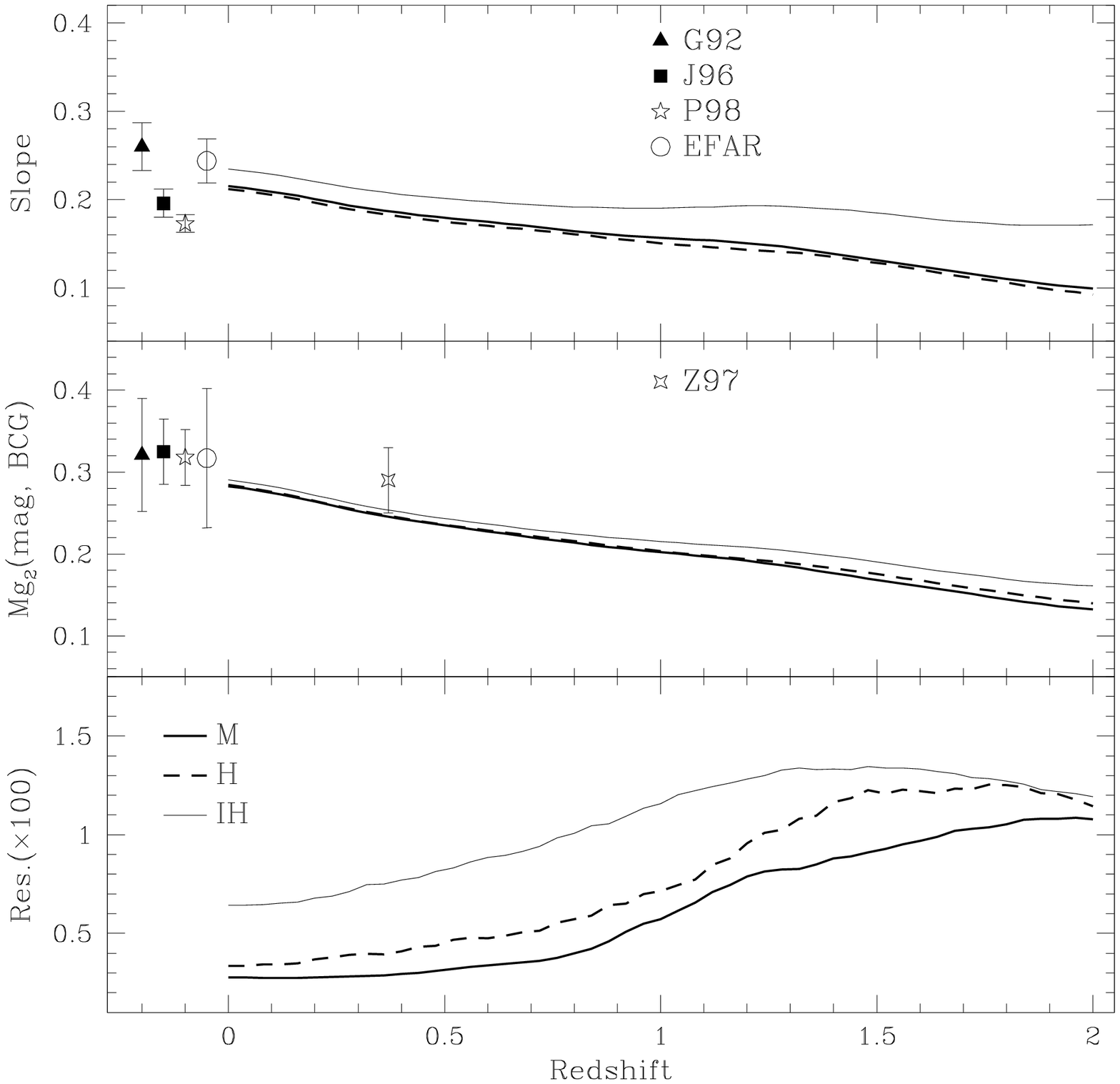}
\figcaption[fs_f7.eps]{Evolution of the magnesium abundance with redshift
explored through the correlation between the Lick/IDS $Mg_2$ index 
(Worthey 1994) and $\log\sigma$. Notice the observations are 
systematically above the predicted curves for slope and zero point
caused by the magnesium overabundance problem, which enhances the magnesium
yields in the central parts of the brightest ellipticals, thereby
steepening the slope and increasing the zero point. The data points come from
observations of local clusters by Guzm\' an et al. (1992, G92);
J\o rgensen et al. (1996, J96); and Colless et al. (1999, EFAR), and 
of a few clusters at $z\sim 0.37$ from Ziegler \& Bender (1997, Z97).
The latter only measure a few galaxies so that it is not possible
to estimate a slope. Therefore the error bar for point Z97 show the range 
in observed magnesium abundances instead.
\label{f7}}
\vskip+0.2truein

%%%%%%%%%%%%%%%%%%%%%%%%%%%%%%%%%%%%%%%%%%%%%%%%%%%%%%%%%%%%%

\noindent 
amount of material (gas, metals and even stars via tidal 
tails) would  be ejected into the intergalactic medium. Furthermore, 
ram pressure stripping from the
hot ICM gas can also play an important role in its enrichment
(e.g. Gunn 1989\markcite{gu89}; Evrard 1991\markcite{ev91}). The origin
of the metal content in the ICM is not only uncertain with respect to which
galaxies contributed the most metals, but the issue of whether most of the metals
came from type~Ia or type~II supernovae is also controversial
(see Ishimaru \& Arimoto 1997\markcite{ia97}; Gibson, Loewenstein
\& Mushotzky 1997\markcite{glm97}; Wyse 1997\markcite{wy97}). 
In our model, we use a fixed ratio of type~Ia's with respect to 
type~II's as described in \S2, and relate most of the metals in the 
ICM to ejecta from low mass galaxies. In this section, we will
trace only the iron abundance, hence the calculations are insensitive
to the parameters related to stars with masses below $M\simlt 10M_\odot$.

Once the metals and the gas ejected into the ICM from all of the galaxies
in the simulated cluster are computed, we need to set a value for the
primordial amount of gas in the ICM which is assumed to have zero
metal content. This is done by a least squares fit of the model curves
compared to observed ICM metallicities from the sample of
Mushotzky \& Loewenstein (1997)\markcite{ml97}, which comprises 
21 clusters in the redshift range $0.1<z<0.6$, observed with the 
{\sl ASCA} X-ray satellite. The primordial mass of ICM gas obtained is
roughly 65\% regardless of the enrichment models, and the evolution
with redshift is shown in figure~8. No clear trend is
evident from the data, and the models predict no change in
ICM metallicity at moderate-to-high redshifts, mainly caused by the
strong star formation efficiency required.
The process of chemical enrichment is rather fast, since
it is primarily driven by massive OB stars with lifetimes of order a few
million years. If formation redshifts $z_F\simgt 2$ are assumed 
for the more abundant component of faint galaxies (which contribute the most
to the ICM because of their high ejected fractions), then by $z\sim 1.5$
most of the metals have been released to the intracluster medium, which
results in no further evolution of $Z_{\rm ICM}$. Hence, the lack of 
evolution in the metal abundance of the ICM observed at moderate redshifts 
is consistent with a picture where most of the metals are expelled from 
the population of faint ellipticals. Were recent and big mergers to 
contribute significantly to the metal abundance of the ICM, then a strong 
drop in $Z_{\rm ICM}$ should be expected by $z\sim 1-1.5$. 
Upcoming X-ray observations of clusters with {\sl AXAF} and {\sl XMM} 
will elucidate this point.

%%%%%%%%%%%%%%%%%%%%%%%%%%%%%%%%%%%%%%%%%%%%%%%%%%%%%%%%%%%%%%%%%%%%%%%%%%%%%

\section{Conclusions}
The strong degeneracy between age and metallicity thwarts any attempt
at estimating the star formation history of early-type galaxies. Hence, 
evolutionary studies must include some model in which both age and
metallicity are related. A simple chemical enrichment model has been
described in this paper, where we have tried to reduce the many factors 
occurring in galaxy evolution down to the most basic parameters. 
We believe that infall of gas, outflows from supernova-triggered winds 
and possible variations in the star formation efficiency can collectively 
determine chemical enrichment. The results from our simple models 
demonstrate that outflows can generate the range in metallicity required 
for the observed color-magnitude relation (figure~1), whereas the  
other principal parameters, associated with infall of primordial gas or 
with the 
star formation efficiency, cannot account for the observations. This 
agrees with dynamical-spectrophotometric relations such as the
Faber-Jackson correlation between central velocity dispersion and
absolute luminosity (Faber \& Jackson 1976\markcite{fj76}). The ejected fraction
of gas should be a function of the mass of the galaxy. It could be argued that
there might be a correlation between star formation and galaxy mass so that
stars would be preferentially formed in more massive galaxies. This would
increase the ejected fraction in these galaxies. However, the 
color-magnitude diagram would then be strongly altered, blueing the bright
end of the red sequence. One way out of this would include other mechanisms
``conspiring'' to restore the redness of the brightest galaxies. Therefore,
we conclude that the simplest explanation for the color-magnitude relation
comes from a range of metal abundances caused by the correlation between
galaxy mass and ejected fraction. We  emphasize that
observational evidence concerning this point is still in a primitive 
stage, and thus a significant age spread cannot be ruled out.
However, observations of the Fornax cluster by Kuntschner \& Davies
(1998)\markcite{kd98} seem to point to a metallicity spread as being responsible
for the color-magnitude relation. Other factors such as an IMF with
a variable high mass slope or cutoff or variable stellar yields 
could in principle decrease the effect of outflows on the
resulting stellar populations. However, if such factors were to play a significant 
role in affecting the scatter of the color-magnitude relation, 
the expected  correlation of these
factors with galaxy mass should  have other observable implications.

The combination of these chemical enrichment tracks with the latest 
population synthesis models  from BC99
allows us to 
estimate the spectrophotometric properties of cluster early-type 
galaxies. Using the color-magnitude relation and the luminosity function of 
local clusters as constraints, clusters can be simulated over a wide range 
of redshifts for different input parameters and star formation scenarios.
A strong degeneracy is found when varying infall parameters (the rate,
timescale and delay of infall) or the star formation efficiency. This is
just a mapping of the degenerate enrichment tracks shown in 
figures~1 and 2 with respect to metallicity and $U-V$ color, respectively.
Three different star formation scenarios are also considered: monolithic
collapse, and two hierarchical models with formation redshifts dependent
on the luminosity of the galaxy. The predictions
for these models are not as degenerate at high redshifts $z\simgt 1$, 
especially for age-sensitive observables such as the mass-to-light ratio.
The most remarkable difference between the models can be shown in the predicted
color-magnitude relations in figure~4: hierarchical models (H and IH)
clearly display at high redshift a population of early-type galaxies which
fall conspicuously blueward of the ``main sequence'' red envelope: these
blue outliers are so far the best candidates for characterizing cluster
evolution. A hierarchical scenario (H-model) presents these
outliers as the brightest cluster galaxies, whereas an inverted hier-

%%%%%%%%%%%%%%%%%%%%%%%%%%%%%%%%%%%%%%%%%%%%%%%%%%%%%%%%%%%%%
%%%%%%%%%%%%%%%%%%%%%%%%   FIG. 8   %%%%%%%%%%%%%%%%%%%%%%%%%
%%%%%%%%%%%%%%%%%%%%%%%%%%%%%%%%%%%%%%%%%%%%%%%%%%%%%%%%%%%%%

\centerline{\null}
\vskip3.4truein
\includegraphics{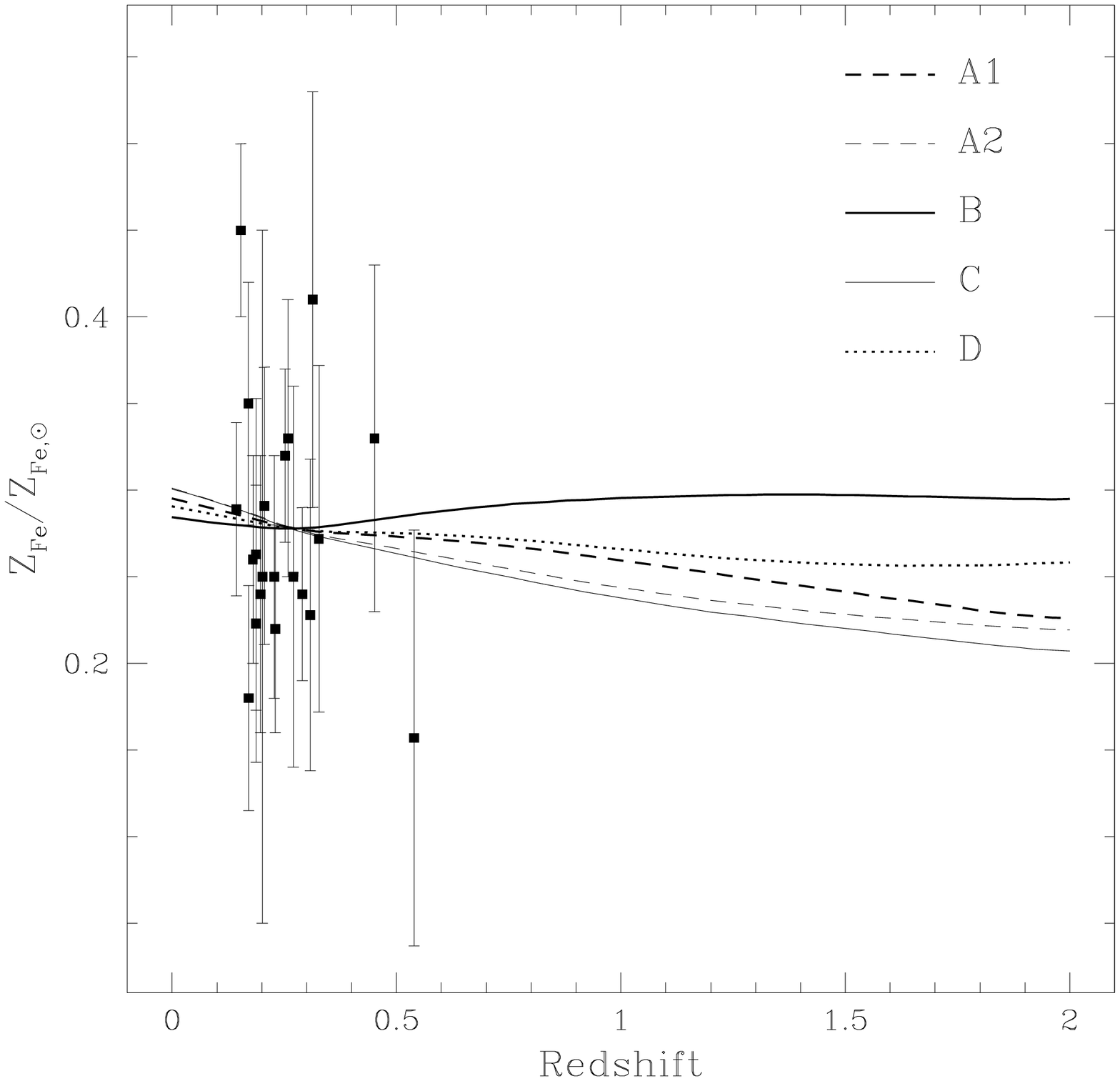}
\figcaption[fs_f8.eps]{Predicted evolution as a function of redshift
of the iron content of the intracluster medium from gas ejected from 
early-type galaxies to a primordial mass of gas which comprises 65\% 
of the present amount in a local rich cluster. The data points come from 
Mushotzky \& Loewenstein (1997). If a significant change in the metallicity 
of the ICM were eventually detected, a very recent and strong star 
formation process should be introduced, since most of the metals are 
produced in massive OB stars with lifetimes of a few 
million years.\label{f8}}
\vskip+0.2truein

%%%%%%%%%%%%%%%%%%%%%%%%%%%%%%%%%%%%%%%%%%%%%%%%%%%%%%%%%%%%%

\noindent
archy (IH-model) identifies them with a faint population of dwarf galaxies. 
Monolithic collapse results in no major difference in the residuals from the 
linear fit with respect to luminosity. 

A complete sample of three blue
outliers down to F814W=21 has been confirmed to be members of 
cluster Cl0016+16 ($z=0.545$) using serendipitous observations of these 
candidates by Belloni \& R\"oser (1996)\markcite{br96} and the MORPHS 
collaboration (Dressler et al. 1999\markcite{dr99}). They are also classified
as post-starburst (E+A or k+a) galaxies 
(Dressler \& Gunn 1983\markcite{dg83}), a clear sign of
recent star formation activity. The observed equivalent widths are matched
against the predictions of population synthesis models and an upper limit to
the age between 0.5 and 2 Gyr is inferred for a range of metallicities 
$0.2 < Z/Z_\odot < 2.5$, with the younger estimates corresponding to the
higher metallicities. This sets an upper limit $z_F\simlt 1$ on the epoch
of formation of these galaxies, favoring a hierarchical clustering scenario
where late merging stages could still be associated with significant
star formation. It is still a matter of debate whether the spectral 
signature of young stars in these galaxies corresponds to a large fraction 
of the galaxy mass.  Such quantitative estimates must be performed in 
order to ascertain whether the star formation process that imprinted the 
E+A signature on these galaxies played a dominant role in accounting for 
the observed stellar populations. The evolution of mass-to-light ratios 
and magnesium abundances can also be explored with these simulated 
clusters. Both give the same degeneracy for a range of toy enrichment 
models, and a significant departure at high redshift ($z\simgt 1$) 
between different star formation scenarios. The comparison between 
stellar $M/L$ ratios and observations (that trace the total mass) 
yield a roughly linear correlation between stellar and
total matter: $M_{\rm Tot}\propto M_{\rm St}^{1.15\pm 0.08}$.

Given that our model uses outflows as the main cause for the range in colors
of cluster early-type galaxies, we can quantify the amount of metals 
ejected into the intracluster medium (ICM). The top panel of figure~3 shows
that a faint galaxy with absolute luminosity $M_V\sim -16$ ejects 85\% 
of its gas. This means that a large number of (dwarf) galaxies contribute
significantly to the enrichment of the ICM. This was computed and an
initial mass of primordial intracluster gas was assumed. The final value
of this initial mass is estimated by comparing the predicted tracks 
$Z_{\rm ICM} = Z_{\rm ICM} (z)$ with cluster X-ray observations from
Mushotzky \& Loewenstein (1997)\markcite{ml97}. A large fraction of the
intracluster medium gas must be primordial ($\sim 65$\%) in order to 
explain these observations. Figure~8 shows that no important evolution 
in the metal abundance of the ICM is expected up to redshifts
$z_F\sim 2$, in rough agreement with the data points. This can be explained
because the fraction of metals contributed from ejecta to the ICM is 
very high even at high redshifts: most of the metals synthesized 
in a galaxy come from young stars with lifetimes of order a few
million years. Subsequently, intermediate mass stars take over the 
enrichment process, but they only contribute to a small fraction of 
the enrichment. Figure~1 shows that any chemical enrichment track using 
passive evolution results in a sharp rise in metallicities followed by a 
stable stage in which these metallicities do not change much. Hence, 
the study of ICM metal abundance in high redshift clusters with 
forthcoming X-ray telescopes such as {\sl Chandra} and {\sl XMM} will 
allow us to estimate the role of enrichment from dwarf galaxies or from 
merger remnants. The latter should translate into a sharp decrease of 
$Z_{\rm ICM}$ at redshifts $z\simgt 1-2$.

\acknowledgments
We would like to thank Sandra Faber for her very useful remarks and 
suggestions about this paper and in general about the eternal 
``tug-of-war'' between observations and theories.
I.F. acknowledges financial support for this project from the Gobierno de 
Cantabria, Institut d'Astrophysique de Paris and the Center for Particle 
Astrophysics at the University of California, Berkeley. J.S. was supported
at Berkeley in part by a grant from NASA.

\end{document}